\documentclass[12pt]{article}
\usepackage{setspace}
\usepackage{etex,mathpazo}
\usepackage{xr}
\usepackage{arydshln}
\usepackage{pdflscape}
\usepackage{amsmath,amsthm,amssymb} %Package para matemáticas.
\usepackage{natbib} %Bibliography
\usepackage[font=normalfont,tableposition=top]{caption}
\usepackage{float}%To improve floating environments
\usepackage{appendix}%To format the appendix.
\usepackage{color}
\usepackage{fullpage} 
\usepackage{chngcntr}

\usepackage{hyperref}
\hypersetup{%
	colorlinks = true,
	linkcolor  = blue,
	citecolor  = blue
}

\makeatletter
\renewcommand*{\eqref}[1]{%
  \hyperref[{#1}]{\textup{\tagform@{\ref*{#1}}}}%
}
\makeatother
\usepackage{bera}
\usepackage{listings}
\usepackage{xcolor}

\colorlet{punct}{red!60!black}
\definecolor{background}{HTML}{EEEEEE}
\definecolor{delim}{RGB}{20,105,176}
\colorlet{numb}{magenta!60!black}

\lstdefinelanguage{json}{
	basicstyle=\normalfont\ttfamily,
	numbers=right,
	numberstyle=\scriptsize,
	stepnumber=1,
	numbersep=2pt,
	showstringspaces=true,
	breaklines=true,
	frame=lines,
	backgroundcolor=\color{background},
	literate=
	*{0}{{{\color{numb}0}}}{1}
	{1}{{{\color{numb}1}}}{1}
	{2}{{{\color{numb}2}}}{1}
	{3}{{{\color{numb}3}}}{1}
	{4}{{{\color{numb}4}}}{1}
	{5}{{{\color{numb}5}}}{1}
	{6}{{{\color{numb}6}}}{1}
	{7}{{{\color{numb}7}}}{1}
	{8}{{{\color{numb}8}}}{1}
	{9}{{{\color{numb}9}}}{1}
	{:}{{{\color{punct}{:}}}}{1}
	{,}{{{\color{punct}{,}}}}{1}
	{\{}{{{\color{delim}{\{}}}}{1}
	{\}}{{{\color{delim}{\}}}}}{1}
	{[}{{{\color{delim}{[}}}}{1}
	{]}{{{\color{delim}{]}}}}{1},
}

\usepackage{chngcntr}
\usepackage{setspace}
\newtheorem*{example cont.}{Example (Cont.)}

\usepackage{bm}

\usepackage{graphicx}
\usepackage{adjustbox}	% this package is for resizing table out of text width

\graphicspath{ {Figures/} }  	% this package is for setting the path of plots folder
\usepackage{graphics}
\usepackage{multirow}

\usepackage{caption}
\usepackage{subcaption}
  \clearpage 
\usepackage{titling}

\usepackage{booktabs}
\newcommand{\ra}[1]{\renewcommand{\arraystretch}{#1}}

\usepackage[left=1in, right=1in, top=1in, bottom=1in]{geometry}
\usepackage{setspace} % load the package to change interline
\usepackage[T1]{fontenc}
\usepackage{caption}
\allowdisplaybreaks[4]

\usepackage{titlesec}
\titlespacing*{\section}{0pt}{0.5\baselineskip}{0.01\baselineskip}
\titlespacing*{\subsection}{0pt}{0.5\baselineskip}{0.01\baselineskip}

\usepackage{colortbl}

\expandafter\def\expandafter\normalsize\expandafter{%
    \normalsize
    \setlength\abovedisplayskip{3pt}
    \setlength\belowdisplayskip{3pt}
    \setlength\abovedisplayshortskip{50pt}
    \setlength\belowdisplayshortskip{50pt}
}

\usepackage{enumitem}

\usepackage{mathtools}
\mathtoolsset{showonlyrefs}

\usepackage[bottom]{footmisc}

\makeatletter
\def\@makefntext{\hskip 0em\@makefnmark}
\makeatother
\begin{document}

%\setstretch{1}

%\begin{titlepage}

\setlength{\droptitle}{-9em}

\title{{\textbf{Inside the Mind of Investors during the COVID-19 Pandemic: Evidence from the StockTwits Data\thanks{Monash Centre for Quantitative Finance and Investment Strategies has been supported by BNP Paribas. I am grateful to the StockTwits company, Garrett Hoffman, for facilitating the access to the data by providing an API. I also thank Yuan Zhang for providing several Python scripts. }}}}
%\renewcommand\footnotemark{}
%\date{ \today}
\date{\vspace{-7ex}}
%Job market paper, please do not cite without permission}
%\date
\author{ 
Hasan Fallahgoul\thanks{Hasan Fallahgoul, Monash University, School of Mathematics and Centre of Quantitative Finance and Investment Strategies, Melbourne, Australia. E-mail: hasan.fallahgoul@monash.edu}
\bigskip
}
\noindent
 \maketitle
\noindent
\vspace{-2cm}
\begin{abstract}
	\noindent
	We study the investor beliefs, sentiment and disagreement, about stock market returns during the COVID-19 pandemic using a large number of messages of investors -- about 3.7 million messages -- on a social media investing platform, \textit{StockTwits}. The rich and multimodal features of StockTwits data allow us to explore the evolution of sentiment and disagreement within and across investors, sectors, and even industries. We find that the sentiment (disagreement) has a sharp decrease (increase) across all investors with any investment philosophy, horizon, and experience between February 19, 2020, and March 23, 2020, where a historical market high followed by a record drop. Surprisingly, these measures have a sharp reverse toward the end of March. However, the performance of these measures across various sectors is heterogeneous. Financial and healthcare sectors are the most pessimistic and optimistic divisions, respectively. 
	
\end{abstract}

\thispagestyle{empty}
% \setcounter{page}{1}
%\end{titlepage}
% \setcounter{page}{1}

%----------------------------------------INTRODUCTION--------------------------------------------------------

\setstretch{2}

%\noindent\textbf{One sentence summary}:

\section{Introduction}
We explore investor beliefs, sentiment and disagreement (dispersion of views across investors), on stock market returns during the COVID-19 pandemic using a large number of messages of investors on a social media investing platform, \textit{StockTwits}. StockTwits is a microblogging platform in which investors can post their views as a Tweeter type format, e.g., 140 characters.\footnote{StockTwits was founded in 2008 as a social networking platform for investors on the financial markets. It is similar to Tweeter with more additional options that are designed for investors to express their views. \cite{cookson2018don} provide an in-depth analysis of the StockTwits data for measuring disagreement and exploring its source. To have reliable and high quality data, they conduct their analysis of messages posted between January 2013 and September 2014. Since then, the number of users, as well as messages, haven exponentially increased. For example, the number of messages posted between December 2019 and March 2020 is about a quarter of their messages. Although we have access to the whole dataset of StockTwits, however, we limit our analysis on posted messages during the COVID-19 pandemic. } Perhaps more importantly, they also can label their messages with a sentiment which can be \textit{bullish}, \textit{bearish}, or leave it unspecified, \textit{neutral}. Since extracting sentiment from text data is a challenging task, this is an important feature.\footnote{A major problem in textual analysis is related to the correct interpretation of context in which certain words are used. In other words, it is difficult to evaluate what truly is a positive, neutral, and negative statement. For example, it is not clear how to deal with sarcasm, see \cite{rosenthal2019semeval}, among others. } Besides, they can use "cashtags", e.g., \$AAPL, to link their message to a particular firm.\footnote{Detailed information about why do people post messages? and why do their messages represent their view? can be found in \cite{cookson2018don} and therein references.}

Using a survey conducted on wealthy retail investors who are clients of Vanguard, \cite{giglio2020inside} provide a data-driven analysis of how investor expectations about economic growth and stock market returns changed during the February-March 2020 stock market crash induced by the COVID-19 pandemic. They find that, among others, investor beliefs before and during the COVID-19 crisis in February-March 2020 shows a sharp increased disagreement about the stock market outcomes and future of the economy. This paper complements their findings in several important ways by analyzing the messages on StockTwits. 

\subsection{Primary contributions}
The contribution of this paper is three-fold. First, thanks to the availability of investors' messages at the high-frequency level, e.g., minute, augmented with their sentiment and cashtags, we can measure disagreement at a high-frequency level.\footnote{The number of posted messages per minute is on the scale of hundreds.} Comparing the disagreement and its evolution which are extracted from StockTwits with their counterparts in \cite{giglio2020inside} is an interesting exercise, given the heterogeneity in the sources of data. 

Second, it is important to measure the disagreement around the COVID-19 pandemic, however, understanding the source of disagreement is more vital for both the design of the ongoing economic policy response and in further advancing economic theories. It is not clear what is the source of disagreement in \cite{giglio2020inside}:  \textit{information} or \textit{interpretation of information}. The StockTwits data can be used to explore this direction of disagreement as in \cite{cookson2018don}.

Third, the rich and multimodal nature of the StockTwits data allows us to take a magnifier on sentiment/disagreement and explore how does it vary across sectors, as well as investors: \textit{homogeneous} or \textit{heterogeneous}. As we have seen while the price of most companies has been dropped during the COVID-19 pandemic, some pharmaceutical and technological companies had a big jump in their price. For example, a tech company such as Zoom had more than $ 20\% $ jump in its price. Hence, these companies might have different sources of disagreement with stocks that dropped. This aspect of StockTwits data can help to understand this direction of disagreement. 

\subsection{Main empirical findings}

To explore the investor beliefs, sentiment and disagreement, on stock market returns during the COVID-19 pandemic, we conduct a comprehensive analysis of all posted messages on the StockTwits platform between November 30, 2020, and March 31, 2020. In total, we have 3,676,169 messages of 179,468 unique users mentioning 10,715 unique tickers. We establish the following empirical facts about sentiment and disagreement of investors that post messages on the StockTwits platform. 

\textit{Daily time series of the sentiment and disagreement is not a stationary process}. This is true across and within all investors with any investment philosophy, horizon, and experience. The same results are valid across sectors. The immediate implication is that before any investigation/analysis it is helpful to deal with non-stationary via either transformation, rolling window, or differencing.

\textit{There is a $ V-$shape ($ \Lambda-$shape ) in sentiment (disagreement) between February 19, 2020 and March 31, 2020.} Specifically, there is a sharp decrease (increase) in sentiment (disagreement) until March 23, 2020, and then they reverse back until the end of our sample. This result is consistent with findings of \cite{giglio2020inside} that the average investor turned more pessimistic about the short-run performance of both stock markets and the economy.

\textit{The pattern of sentiment and disagreement is homogeneous for investors across and within investment philosophy, horizons, and experiences}. Daily time series of sentiment and disagreement support this finding. Furthermore, the average and standard deviation of sentiment and/or disagreement are ranging from $ 0.6 $ to $ 0.8 $ during the COVID-19 pandemic. 

%\textit{The differential interpretation of information has likely a small contribution as a source of disagreement during the COVID-19 pandemic}.\footnote{We believe a deeper analysis is required to support this finding. One direction to tackle this issue is to replicate the analysis of \cite{cookson2018don} with our dataset. } We calculate the disagreement correlation matrix within investment philosophy, horizon, and experience, and find that there is a strong correlation in the disagreement within these group of investors.

\textit{The pattern of sentiment and disagreement is heterogeneous across sectors.} Daily time series of sentiment and disagreement across sectors behave quite differently across sectors which is in contrast with the behavior of sentiment and disagreement across investors. For instance, the daily time series of sentiment for the healthcare sector is almost flat while it has a downward trend for the financial sector.

\textit{The financial sector is the most pessimistic while healthcare is the most optimistic during the COVID-19 pandemic.} The median of sentiment for the financial sector is $ 0.257 $ while this statistic for the healthcare sector is $ 0.782 $. Moreover, the disagreement for the financial sector is $ 0.901 $ which is pretty high while this statistic for disagreement of healthcare is $ 0.581 $.

%\subsection{Related literature}

\section{StockTwits}\label{sec:stocktwits}

To explore investor beliefs, sentiment and disagreement, on stock market returns during COVID-19 pandemic, we limit our sample on messages posted between November 30, 2019 and March 31, 2020.\footnote{The reason for this range is the following. It is believed that the virus has started spreading in Wuhan in the beginning of December 2019. Also, we are only able to update our dataset once per month, at the end of each month. The last month in our dataset at the conducing this paper was March 2020. } The format of downloaded messages is JSON.\footnote{See Listing \ref{listing_json_file} for a detailed information about JSON format and an example of it from StockTwits. } We use the Python 3.7 for our analysis.

In total, we have 3,676,169 messages of 179,468 unique users mentioning 10,715 unique tickers. For each message/idea, we observe a unique message and user identifier among much other information, see Listing \ref{listing_json_file}. Panel A of Table  \ref{tab_characteristics_message_users} presents summary statistics for some characteristics of raw StockTwits Data, without any filtration. The average number of messages per stock-date is 23.99, with as many as 26,424 messages for some stocks on some days. The average number of people a user follows is 23.92 where this stat for the number of followers is 121.37. The median number of ideas (likes) a user has is 37 (22), with as many as 2,126,704 (809,351) ideas (likes).

Textual analysis of a dataset by Natural Language Processing (NLP) techniques requires an initial assessment. We do need an NLP technique for extracting the sentiment of unspecified messages where they are neither bullish nor bearish. Panel B of Table \ref{tab_characteristics_message_users} shows the frequency distribution of messages' basic features such as the number of words, characters, stopdwrods, and cashtags, as well as, the average length of words. $ \$SPY $ is the cashtage with the highest number of frequencies which is 368,086 (10\% of total cashtags). 

\begin{figure}[!ht]
	\caption{Frequency of cashtags, industries, and sectors.}
	\caption*{Top: this figure presents frequency of distribution of castages during COVID-19 pandemic on StockTwits platform. Middle: this figure presents frequency of distribution of messages of top 10 industries. Bottom: this figure shows frequency distribution of messages by sector. ETF: Exchange Traded Fund. BioTech: Biotechnology. Auto Manu.: Auto Manufacturers - Major. Elc Util.: Electric Utilities. Aerospace: Aerospace/Defense Products \& Services. SemiCond.: Semiconductor - Broad Line. Close-End: Closed-End Fund - Debt. Oil \& Gas: Independent Oil \& Gas. Pers. Comp.: Personal Computers. Med. Lab.: Medical Laboratories \& Research.
	}
	\renewcommand{\thefigure}{\arabic{figure}}
	\begin{subfigure}[b]{1\textwidth}
		\centering
		\includegraphics[scale=0.25]{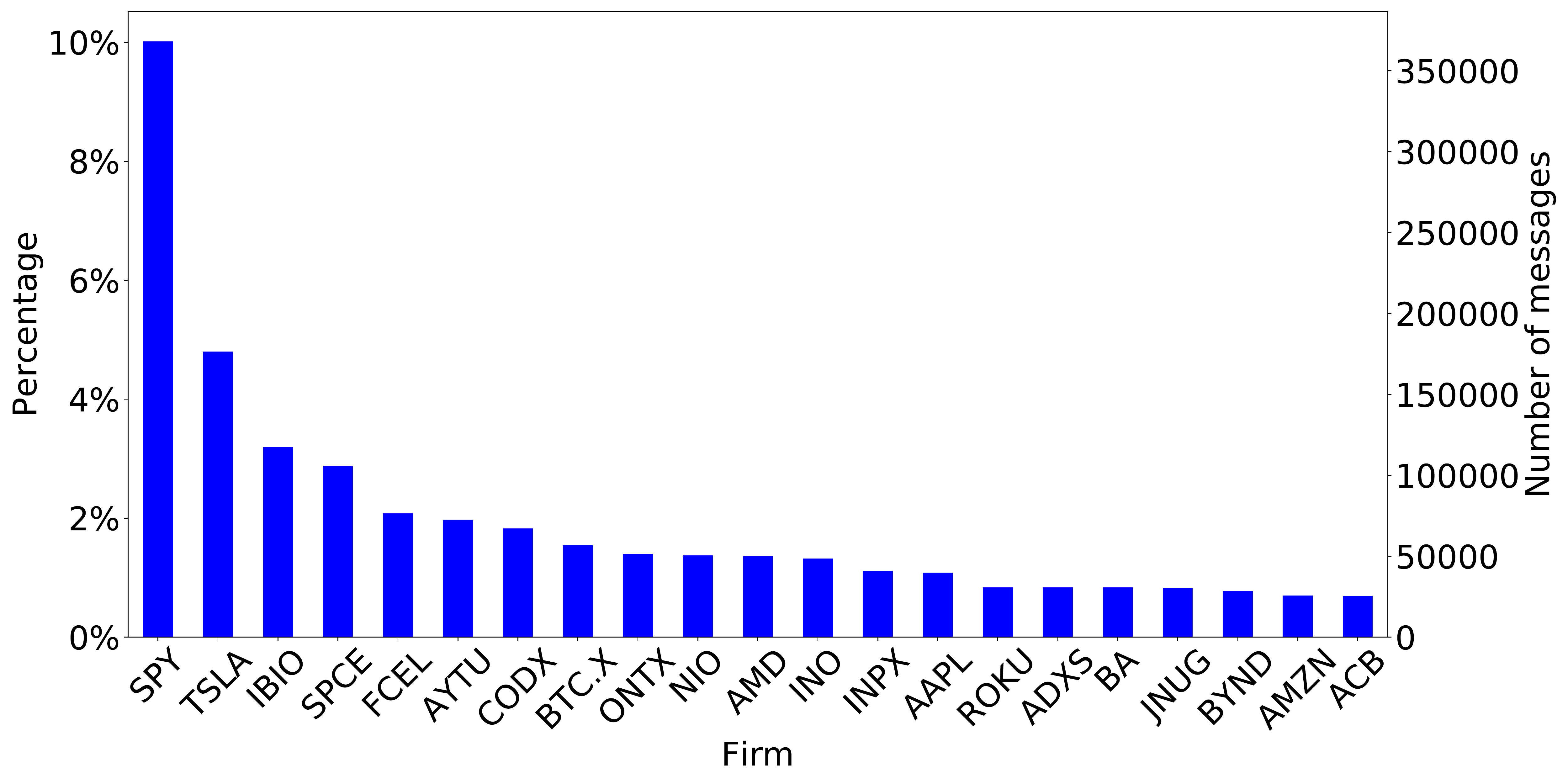}                               
		%             \caption{Day-of-week frequency distribution of messages posted to StockTwits}
		%        \label{top10_repeared_wored}
	\end{subfigure} 
	\begin{subfigure}[b]{1\textwidth}
		\centering
		\includegraphics[scale=0.25]{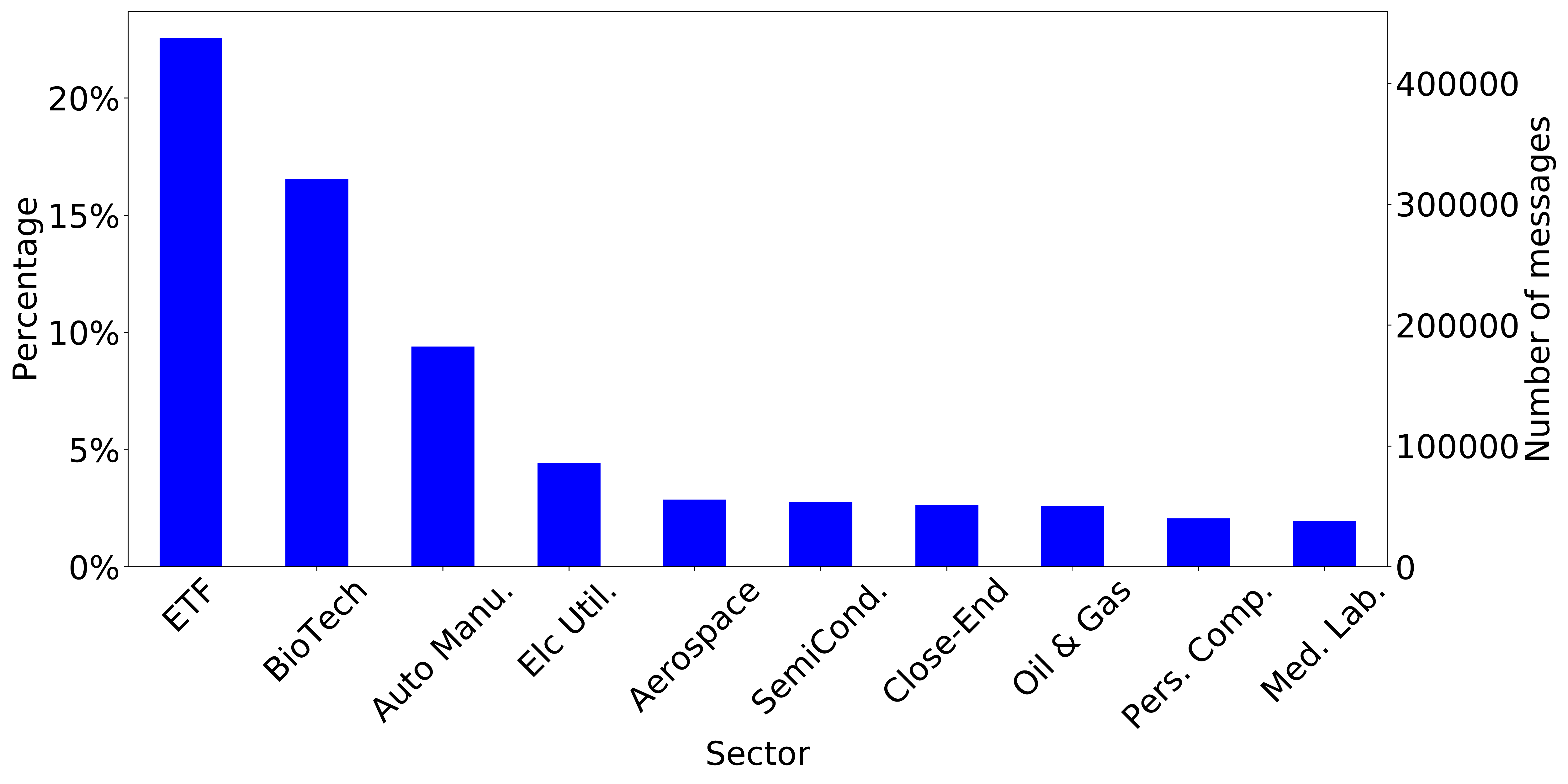}                               
		%             \caption{Day-of-week frequency distribution of messages posted to StockTwits}
		%        \label{top10_repeared_wored}
	\end{subfigure} 
	\begin{subfigure}[b]{1\textwidth}
		\centering
		\includegraphics[scale=0.25]{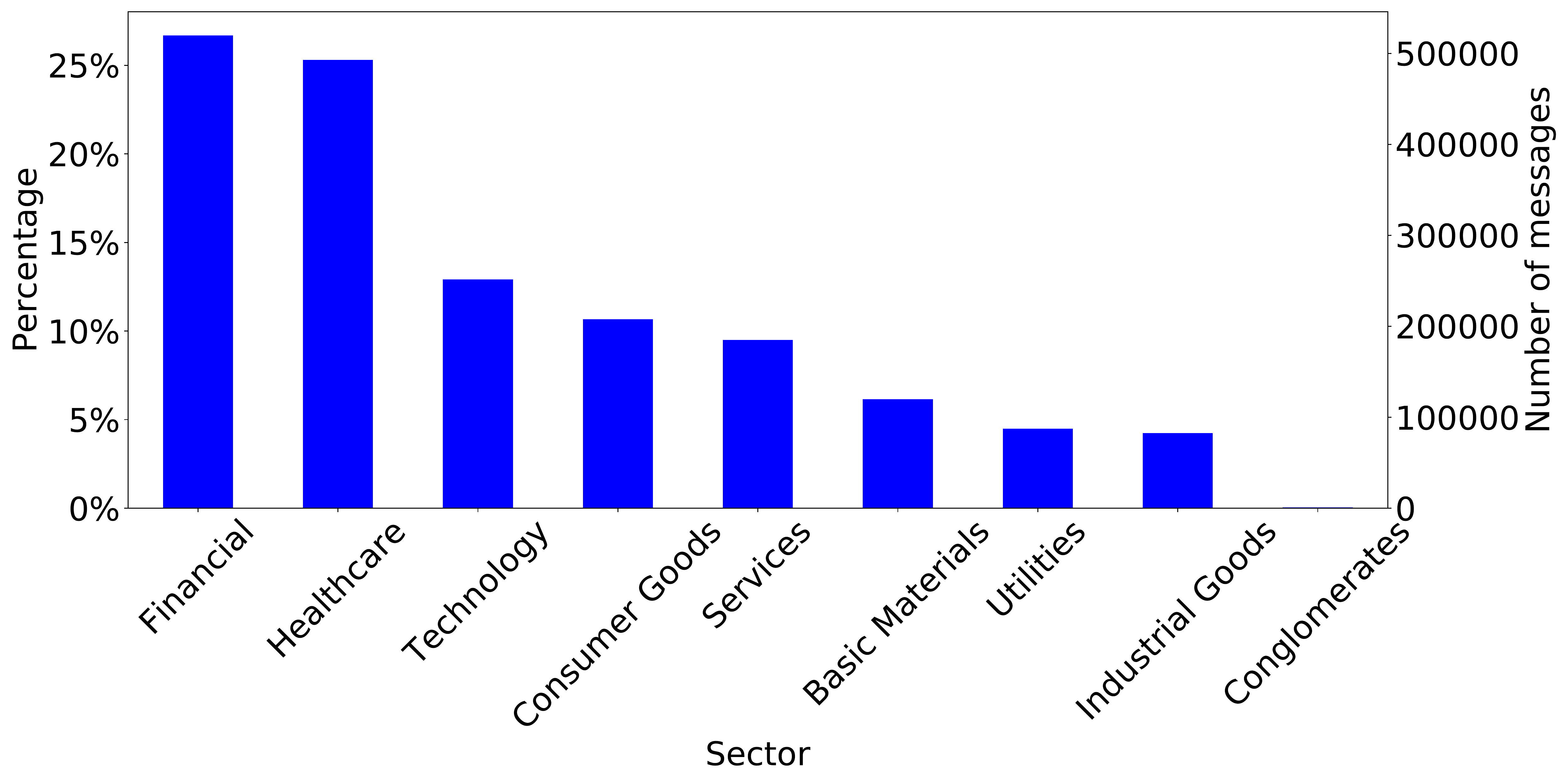}                               
		%             \caption{Day-of-week frequency distribution of messages posted to StockTwits}
		%        \label{top10_repeared_wored}
	\end{subfigure} 
	\label{fig_top20_cashtages_sector}
\end{figure}

The top graph of Figure \ref{fig_top20_cashtages_sector} exhibits the ticker of most frequent firms, as well as, their frequency during the COVID-19 pandemic. The most mentioned firm is Tesla with 176,540 times (4.8\% of total cashtags) while in the bottom of the top 20 firms there are companies such as Amazon and Aurora Cannabis Inc. with the frequency of 25,767 and 25,502, respectively. The middle graph of Figure \ref{fig_top20_cashtages_sector} reveals top 10 sectors with most frequency. Exchange traded fund are most mentioned industries with more 22.55\% of posted messages. Medical Laboratories \& Research industry are 10th with 1.9\% of posted messages. The bottom graph of Figure \ref{fig_top20_cashtages_sector} shows the frequency distribution of posted messages by sector during this pandemic. \cite{cookson2018don} find that the technology and pharmaceutical companies are the most frequent. However, during this pandemic financial and healthcare sectors have the most frequencies.  

\begin{figure}[!ht]
	\caption{Hour-of-day and day-of-week frequency distribution of messages posted to StockTwits.}
	\caption*{This figure presents a frequency distribution of messages posted by  hour of the day (Eastern Standard Time; top) and day of the week (bottom) that messages are posted to StockTwits. Trading hours (days) are plotted as blue bars and nontrading hours (days) are plotted as red bars. }
  \renewcommand{\thefigure}{\arabic{figure}}
  	\begin{subfigure}[b]{1\textwidth}
	\centering
  	\includegraphics[scale=0.9]{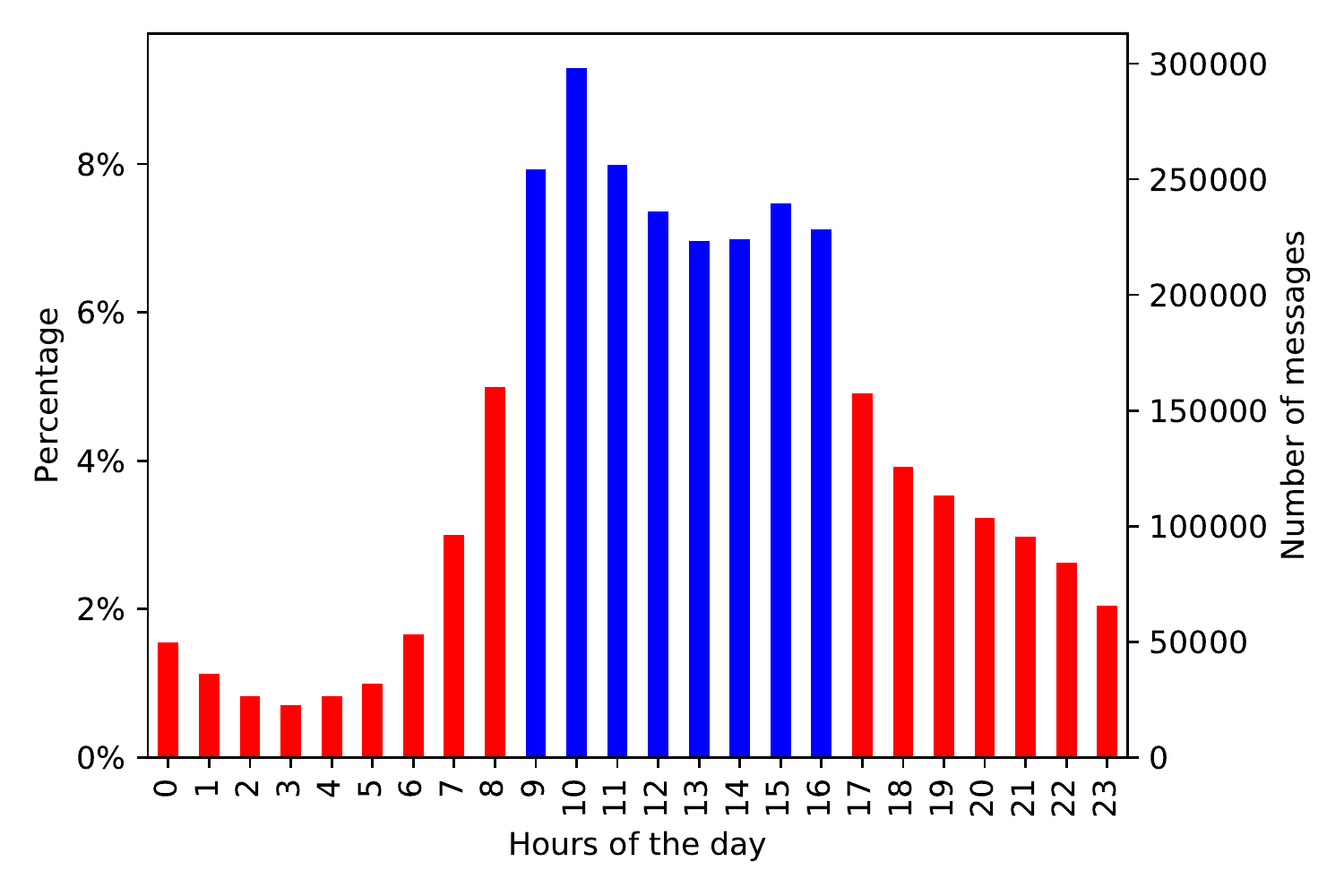}                               
%             \caption{Day-of-week frequency distribution of messages posted to StockTwits}
%        \label{top10_repeared_wored}
    \end{subfigure} 

	\begin{subfigure}[b]{1\textwidth}
		\centering
		\includegraphics[scale=0.9]{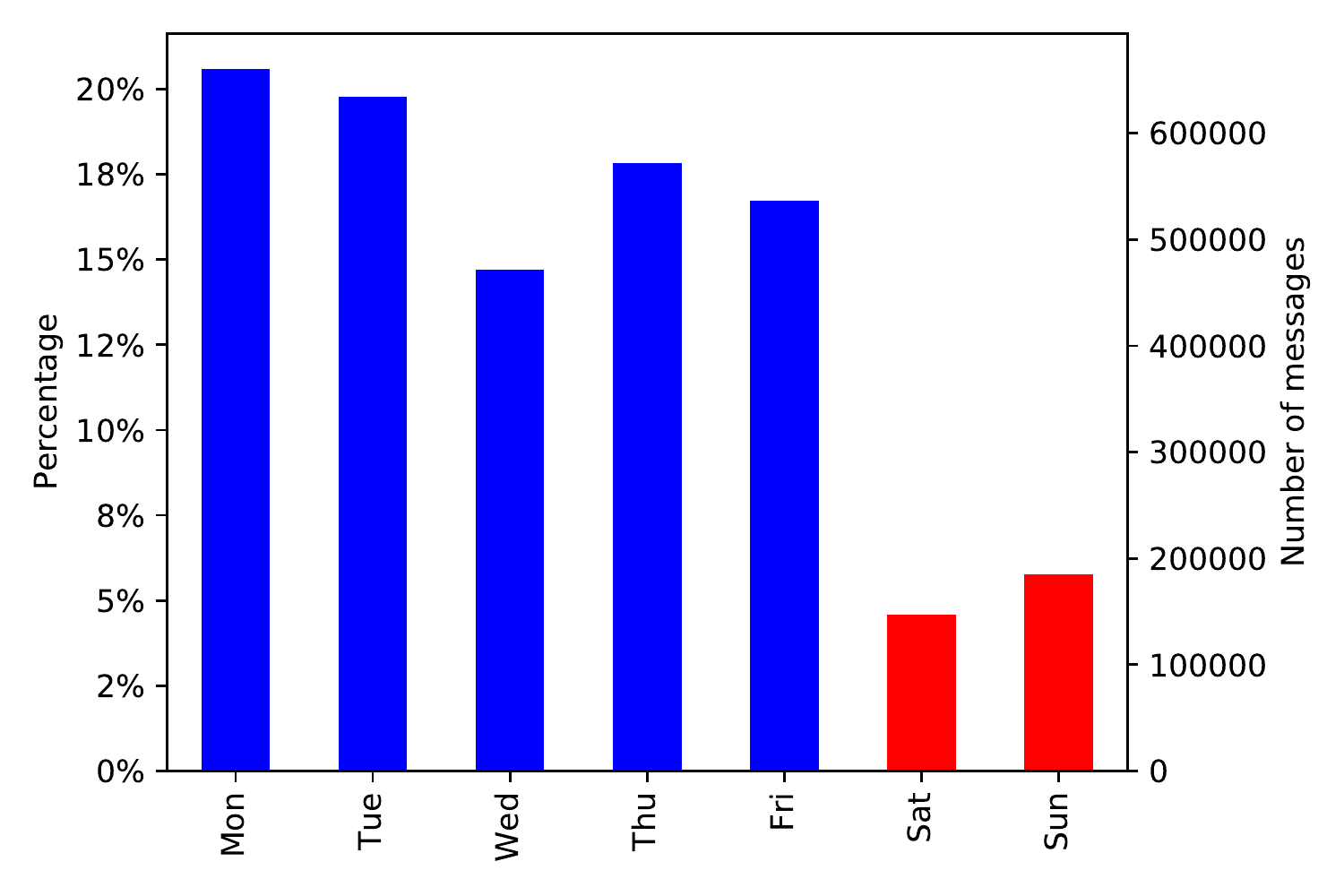}                               
%		\caption{bla bla bla ...}
%		\label{top10_repeared_wored_percentage}
	\end{subfigure} 
\label{fig_freqency_hour_day}
\end{figure}

Figure \ref{fig_freqency_hour_day} exhibits the distribution of messages by the day of the week and by the hour of the day. As can be seen, investors tend to post messages during the opening of the market (Monday-Friday, between 9 a.m. and 4 p.m.). As pointed out by \cite{cookson2018don}, this timing is consistent with investors updating their messages in real time as financial events unfold.

\subsection{Investor philosophies}

StockTwits's users can fill out their profiles with information about themselves as investors. Specifically, they are able to specify their investment approach, investment horizon (holding period), and experience level. Table \ref{table:user_message_approach}, we present the breakdown of users by investment approach, holding period, and experience. This table reveals several important messages. First, more than 80\% of users do not specify their approach, investment horizon, as well as, experience level. This is consist with the dataset of \cite{cookson2018don}. The number of users in their dataset drop to 12,029 from 107,808 users after removing those people that do not report their investment approach, holding period, and experience in their profile information. 

Second, on StockTwits, the most common approach is technical, representing 5.92\% of users and about 9.7\% of messages. Growth and momentum investors represent the next two most common investment philosophies (3.44\% and 3.42\% of investors, respectively), followed by fundamental and value investors  (2.48\% and 1.86\% of investors, respectively), representing 3.57\% and 2.6\% of messages, respectively. Global macro investors make up only 0.78\% of overall investors and 0.56\% of messages.\footnote{The same systematic analysis as been done in \cite{cookson2018don} can be conducted to examine whether the StockTwits investment approaches reliably categorize users into truly different investment philosophies.}

Third, on StockTwits, swing trader has the largest number of users for investment horizon, representing 6.85\% of users and 11.78\% messages. Long term investor and day trader are the next two common investment preferences for holding period (4.51\% and 3.61\% of investors, respectively), followed by position traders with 2.98\% and 4.95\% of investors and messages, respectively.

Finally, users with intermediate experience are the most common investors, representing 8.46\% of users and 15.79\% of messages. Novice and professional users share the same percentage of users by 4.92\% and 4.72\% of investors, respectively. However, professional investors tend to post messages more often than novice investors. The percentage of posted messages for professional investors is 8.73\% while this number is 5.10\% for novice investors.

\section{Sentiment and disagreement }

Eliciting sentiment from a text document such as a tweet, message, or review of a product on Amazon, among others, is a challenging but important task. This task means identifying whether a piece of text, e.g., tweet, express positive, negative, or neutral sentiment.\footnote{\cite{rosenthal2019semeval} describe some details about sentiment analysis for tweets.  \cite{gentzkow2019text} provide a gentle introduction to the use of text as an input to economic research. } Thanks to an outstanding feature of StockTwits, allowing users to label their messages, identifying the sentiment of its messages is not a problem. This critical feature enormously facilitates the analysis of StockTwits's messages. Although the field of textual analysis has been provided with many advanced algorithms, however, they can not be as accurate as of the person that posts the message. 

To carry our analysis, we drop the messages that left their sentiment as null. We also eliminate messages with multiple Cashtages from our analysis.\footnote{More then half of the messages on our dataset are unclassified, neither bullish nor bearish. Although, by excluding unclassified messages, we lose about half of our dataset, however, the number of messages that remain is still substantial. Possibly, it would be interesting if we label unclassified messages from classifies ones as it is done in \cite{cookson2018don}. We leave it for further investigation.}  We follow the same approach as \cite{antweiler2004all} and \cite{cookson2018don} in constructing a sentiment measure from bullish and bearish messages. Specifically, we first label each bearish message as $ -1 $ and each bullish message as $ 1 $. We then take the arithmetic average of these classifications at the $ group1\times day\times group2 $ level: 
\begin{align}
 \text{AvgSentiment}_{itg}= \dfrac{N^{Bullish}_{itg} - N^{bearish}_{itg}}{N^{Bullish}_{itg} + N^{bearish}_{itg}}
\end{align}
where $ N^{Bullish}_{itg}  $ and $ N^{bearish}_{itg} $ are number of bullish and bearish messages per group1, day, and group2, respectively. Group1 can be either all firms, sectors, industries, or a specific firm, sector, or industry. Group2 can either be all investors or investors with a given investment philosophy, experience, or holding period (investment horizon) level. 

This measure of sentiment has important features. Among others, measuring the sentiment at the sector or industry level can be potentially very useful. For example, during the COVID-19 pandemic knowing the sentiment at the level of industry can be fruitful for the designing ongoing economic policy response. Unfortunately, this direction of a sentiment measure is not explored in \cite{cookson2018don}.

 Disagreement and sentiment go hand-in-hand and they cannot be considered separately from each other. We calculate the disagreement similar to \cite{cookson2018don} ( \cite{antweiler2004all}) where disagreement is measured by computing the standard deviation of expressed sentiment across $ group1\times day\times group2 $ (messages) as following
\begin{align}
\text{Disagreement}_{itg}= \sqrt{1-\text{AvgSentiment}_{itg}^2}.
\end{align}
where group1 can be either all firms, sectors, industries, or a specific firm, sector, or industry and group2 can either be all investors or investors with a given investment philosophy, experience, or holding period (investment horizon) level. \footnote{Detailed information about this measure can be found in \cite{antweiler2004all} and \cite{cookson2018don}.}

\section{Sentiment dynamics: moving average}
The rich and multimodal features of the StockTwits data enable us to construct a time series of sentiment at the daily frequency. In this section, we discuss possible dynamics for modeling the time series of average sentiment and its decomposition.

\begin{figure}[!ht]
	\centering
	\caption{Daily time series of sentiment and disagreement, as well as, disagreement correlation matrix.}
	\caption*{Panel (a): this figure exhibits the daily time series of sentiment and disagreement of users based on their approach. A user can fill out his/her profile by approach, experience, and holding period (investment horizon). We first label each bearish message as $ -1 $ and each bullish message as $ 1 $. We then take the arithmetic average of these classifications at the $ group1\times day\times group2 $ level: $ \text{AvgSentiment}_{itg}= \dfrac{N^{Bullish}_{itg} - N^{bearish}_{itg}}{N^{Bullish}_{itg} + N^{bearish}_{itg}}$. Group1 can be either all firms, sectors, industries, or specific firm, sector, or industry. Group2 can either be all investors or investors with a given investment philosophy, experience, or holding period (investment horizon) level. Furthermore, to follow the sentiment's trend, we use rolling windows of the past seven days $ \text{AvgSentiment}_{itg} $ for measuring the sentiment at each day. Disagreement is calculated by: $\text{Disagreement}_{itg}= \sqrt{1-\text{AvgSentiment}_{itg}^2} $. Panel (b): these two heatmaps present the disagreement correlation matrix of users by investment philosophy, experience, and holding period. }
	\renewcommand{\thefigure}{\arabic{figure}}
	\begin{subfigure}[b]{1\textwidth}
		\centering
		\includegraphics[scale=0.16]{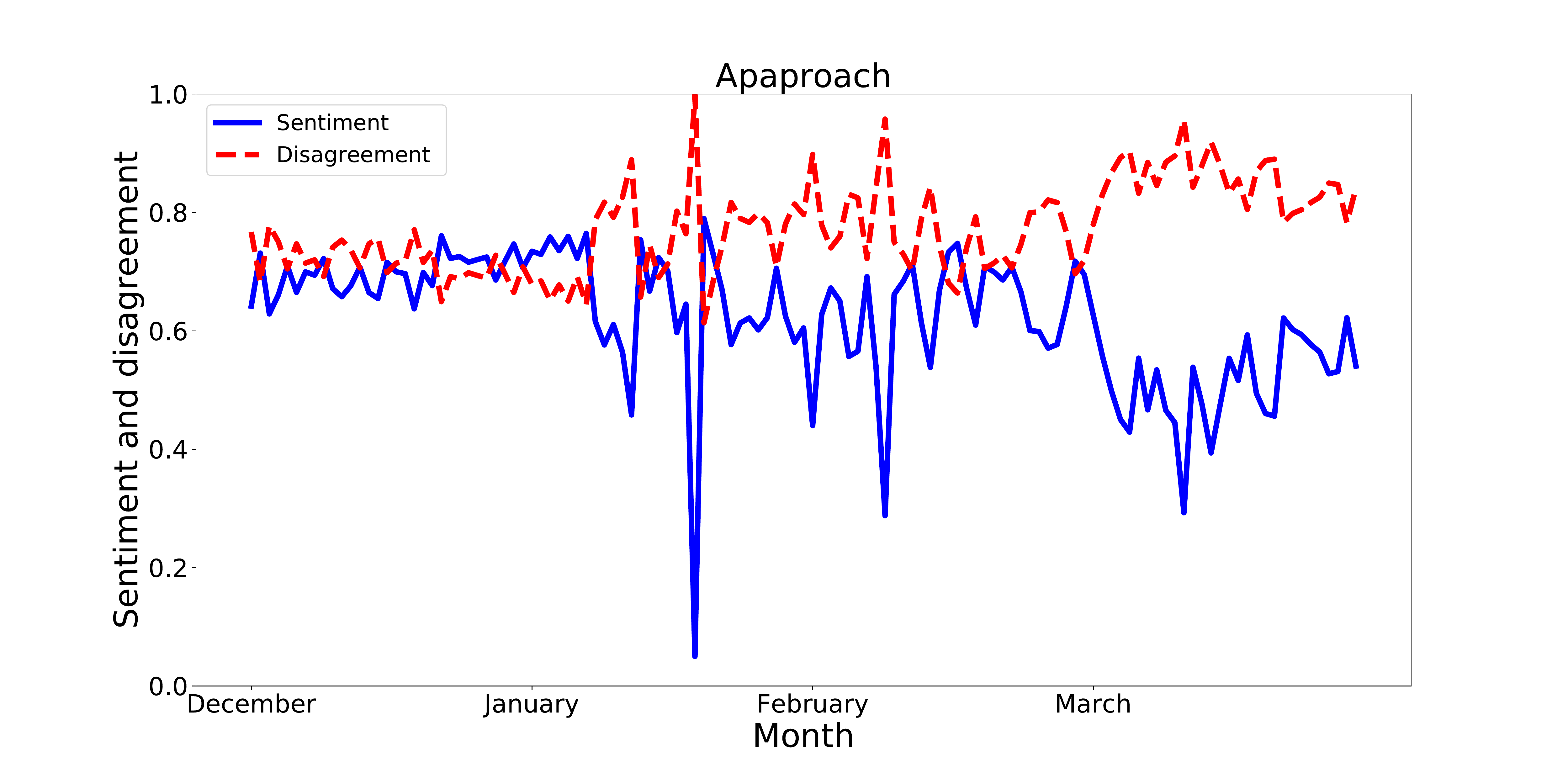} 
		\includegraphics[scale=0.16]{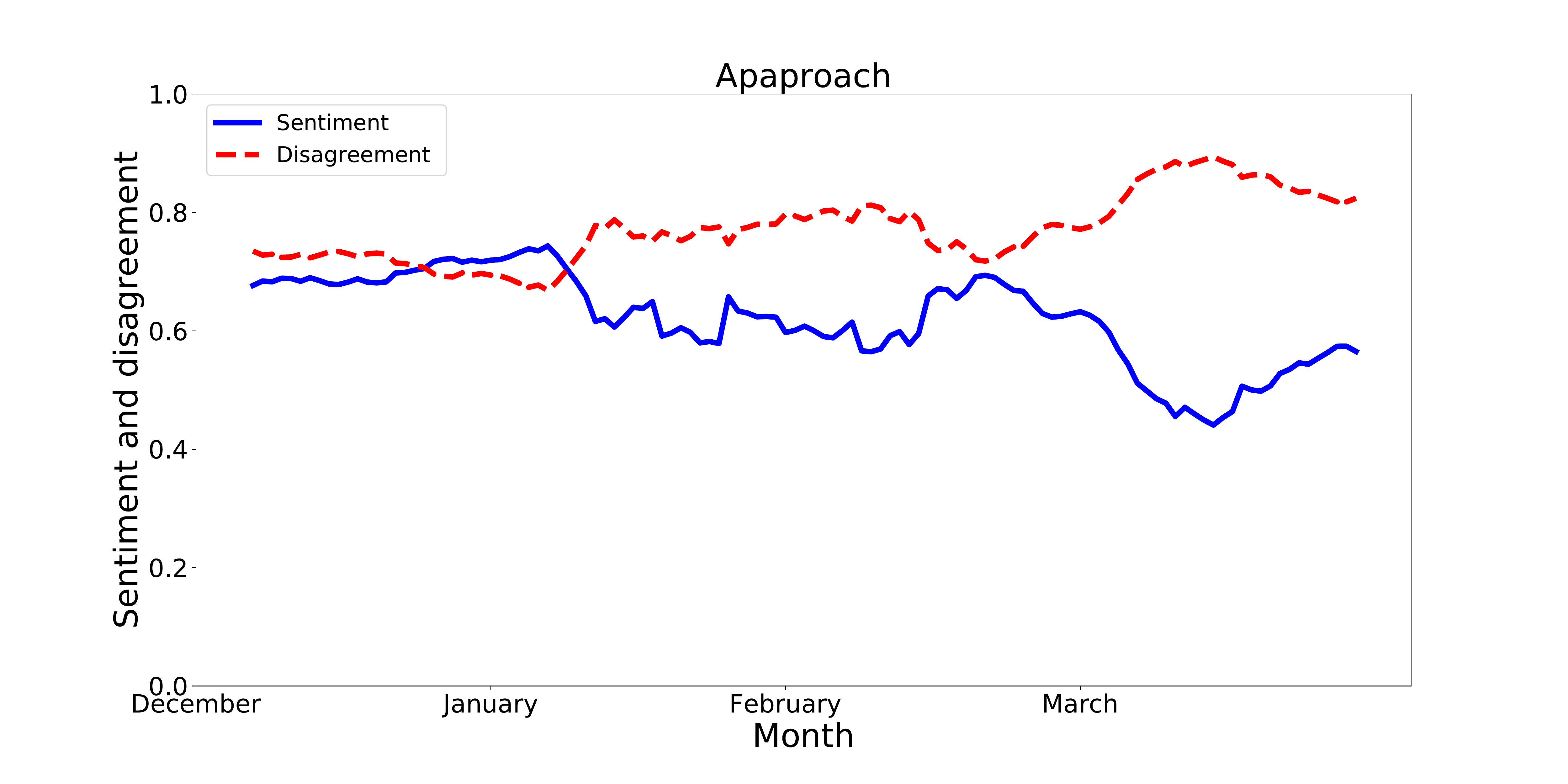}                               
		             \caption{Time series of sentiment and disagreement with (right)  and without (left) rolling window}
		%        \label{top10_repeared_wored}
	\end{subfigure} 
	
	\begin{subfigure}[b]{1\textwidth}
		\centering
		\includegraphics[scale=0.32]{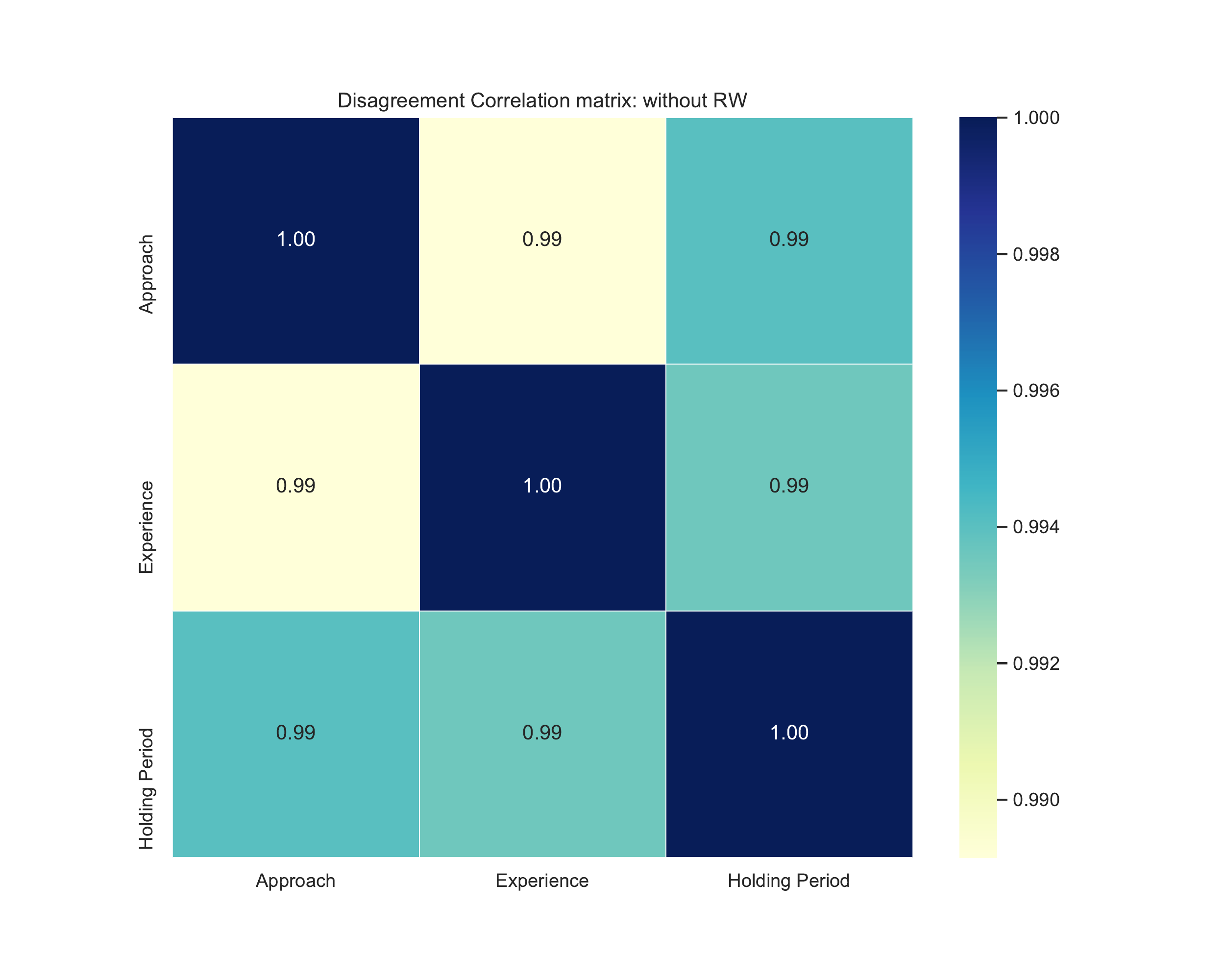} 
			\includegraphics[scale=0.32]{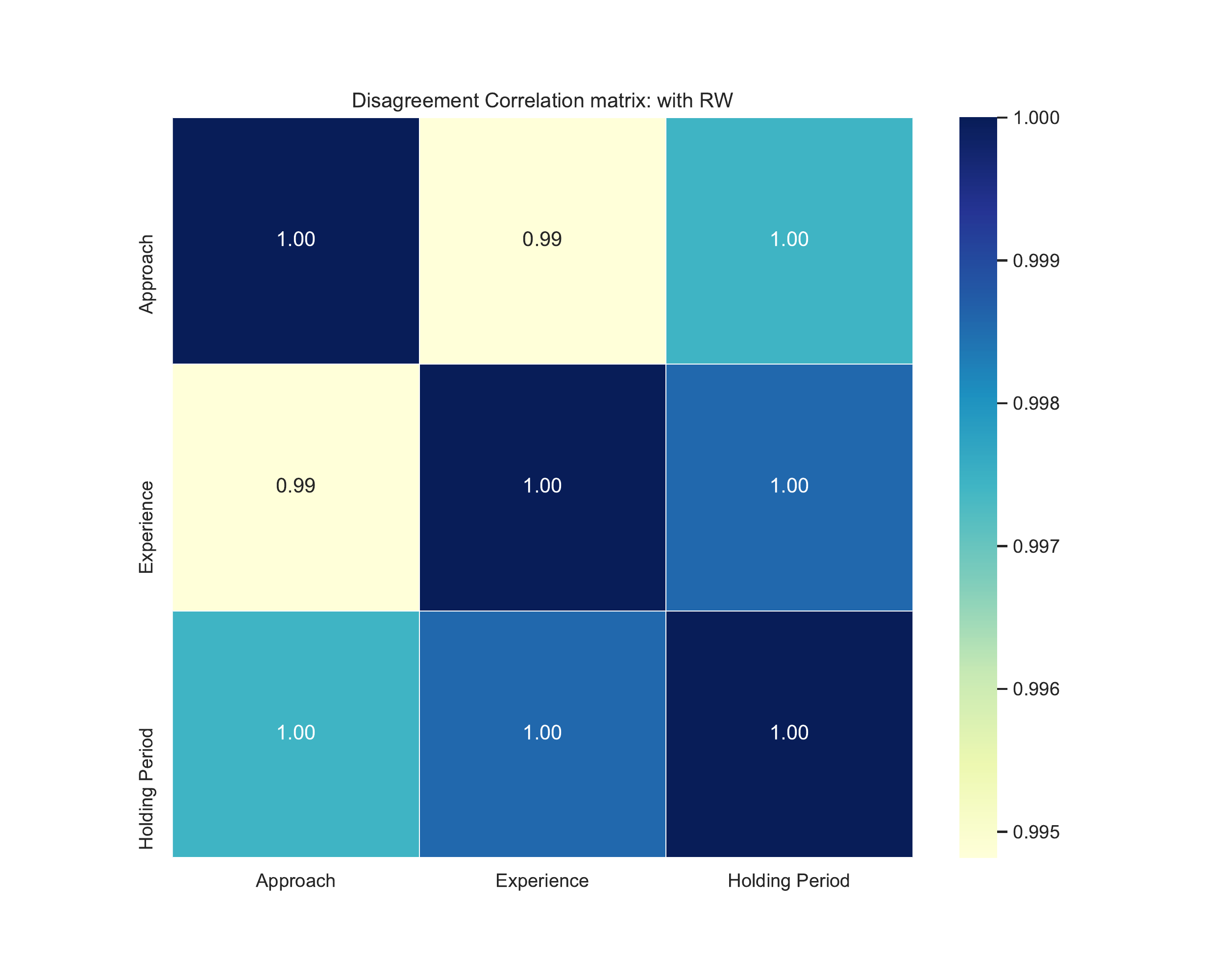}                              
				\caption{Disagreement correlation matrix with (right)  and without (left) rolling window }
		%		\label{top10_repeared_wored_percentage}
	\end{subfigure} 

%\begin{subfigure}[b]{1\textwidth}
%	\centering
%	\includegraphics[scale=0.25]{Fig_disagreement_app_exp_hp_RW.pdf}                               
%	%		\caption{bla bla bla ...}
%	%		\label{top10_repeared_wored_percentage}
%\end{subfigure} 

	\label{fig_sentiment_}
\end{figure}

We calculate the average sentiment measure, $ \text{AvgSentiment}_{itg} $, for day $ t $ from messages posted between the market close of day $ t-1 $ to the market close of day $ t $.\footnote{As in \cite{cookson2018don}, for this analysis, we compute the average sentiment measure by assigning each message an equal weight. However, our dataset allows us to scale the sentiment of each message by the number of likes or number of followers its user has. We leave this robustness check for further investigation.} Panel (a) of Figure \ref{fig_sentiment_} presents the daily time series of sentiment where $ group1 $ is all sectors, $ group2 $ users that group by their investment approach and $ t $ represents a day. As we can see, the time series is noisy. A closer visual inspection on it reveals, this time series is not stationary, that is the mean, variance, and covariance are time-dependent. In this paper, we are interested in the trend of sentiment and disagreement process. Therefore, we need to transform the data somehow to identify its short-term trend. 

Investors often use moving average approach to detect a signal for taking long or short positions, trend-following approach. The sentiment at day $ t $ equals the averages of past seven days of $ \text{AvgSentiment}_{itg} $. We use simple moving average (SMA) which is unweighted mean of previous days. Given the length of our sample, we use the seven days moving average -- one week. However, weighted/exponential moving average with different lags can be considered for this analysis. Specifically, 
\begin{align}
\text{AvgSentiment}_{itg} = \dfrac{1}{7}\sum_{q=1}^{7}  \text{AvgSentiment}_{i(t-q)g}.
\end{align}

The rolling window can be justified as follows. It is rational to assume that the sentiment of a user is unlikely to change instantaneously, contrary, its evolution is likely gradual. Statistically speaking, the process of the sentiment is unlikely to be stationary. Hence, using the past seven observations for extracting the sentiment can be justified.\footnote{There are different directions of this procedure which need to be robustly checked. First, it is important to check the length of the rolling window, the number of past days. Second, what is the right weight for each past observation? In the current rolling window procedure, we give equal weights to all past seven sentiments.} Right side of Panel (a) in Figure \ref{fig_sentiment_} shows the daily time series of sentiment with the rolling window on past seven days. As we can see, the daily time series of sentiment after applying the rolling window is less noisy, perhaps, more importantly, its short-term trend is clear.

In general, depending on the nature of the trend and seasonality, there are two main approaches, additive or multiplicative, wherein, each observation in the series can be manifested as either a sum or a product of the components, respectively. For example, an additive model can have a decomposable time series alike \cite{harvey1990estimation}.\footnote{There is a useful Python and R package for empirical implementation of this model, see \cite{taylor2018forecasting}.} The model has three main components: trend, seasonality, and holidays:
\begin{align}
\text{AvgSentiment}_{itg} = g_t+s_t+h_t+ \epsilon_t
\end{align}
where $ g_t $ is the trend function, $ s_t $ represents periodic changes, e.g., weekly and monthly, $ h_t $ represents the role of holidays, and $ \epsilon_t $ is any idiosyncratic changes which are not accommodated by the model. This specification is similar to a generalized additive model of \cite{hastie1987generalized}.

Panel (b) of Figure \ref{fig_sentiment_} presents the disagreement correlation matrix with (right) and without (left) rolling window where $ group2 $ can be either investment approach, experience, or holding period. \footnote{ To save space, we have not reported the daily time series as Panel (a) for investors' experience and their investment horizon. Their time series have a similar pattern as their counterpart, investment approach.} A close inspection reveals two important messages. First, the rolling window does not affect the structure of the disagreement correlation matrix. The graphs on the left and right side of Panel (b) are very similar. Second, the disagreement is strongly correlated together across investors with any approach, experience, and investment horizon.

%
%%\noindent \textit{Decomposition.}
%
%
%\footnote{We refer reader to \cite{taylor2018forecasting} for details on models of trend, seasonality, and holiday/event.}
%
%
%\noindent \textit{Stationary or non-stationary.} Stationary is an important concept and plays a central role in the modeling of time series (a realization of a stochastic process). Intuitively, it means that the statistical properties of time series process do not change over time.\footnote{We emphasis that the stationary is a property of a stochastic process not a time series.} 
%

%\section{Self- and cross-excitation about sentiment}
%
%\subsection{Modeling contagion: multivariate Hawkes processes}
%
%\subsection{Who influences  who?}
%
%\subsection{What influences what?}
%
%\section{Self- and cross-excitation about industries portfolios}

\section{Empirical findings}

In this section, we explore the evolution of sentiment and disagreement across and within all investors and sectors.

\subsection{Investors}

\textit{Summary}. We find that the sentiment (disagreement) has a sharp decrease (increase) between February 19, 2020, and March 23, 2020, across all investors with any investment philosophy, horizon, and experience, where a historical market hight followed by a record drop. The pattern of sentiment and disagreement is homogeneous for investors with different investment philosophies, horizons, and experiences. Furthermore, The differential interpretation of information has likely a small contribution as a source of disagreement during the COVID-19 pandemic. 

Table \ref{stats_sentiment_user} presents summary statistics on average sentiment broken down by investment philosophy (approach), horizon (holding period), and experience. As investors tend to post bullish messages more frequently than bearish messages, it makes sense that the average sentiment across investors with any investment philosophy, horizon, and experience is closer to 1 than $ -1 $. During our sample period, values investors are the most likely to post bullish messages. Surprisingly, during the sample of \cite{cookson2018don}, technical investors are more likely to post bullish messages. The averages of sentiment for a value investor is 0.721, whereas fundamental, technical, momentum, and growth investors post bearish messages with the same probability, about $ 0.6 $. As for investment horizon, a day trader has the lowest probability to post bullish messages which is 0.372. Furthermore, the standard deviation of the sentiment within day trader investors is the highest, at about $ 0.873$. The probability of posting bullish messages for swing traders, long term investors, and position trader are similar and about $ 0.6 $. Finally, professional investors are less likely to post bullish messages, furthermore, the standard deviation within professional investors is the highest, about $ 0.84 $ whereas this statistic is $ 0.611 $ and $ 0.616 $ for novice and intermediate investors, respectively.

To save space, we do not report daily time series for average sentiment and disagreement within investors' philosophy, experience, and investment horizon.\footnote{These results are available upon request.} We observe the following messages from these time series. Within investors' philosophy, all investors, except growth, share the same opinion on sentiment and disagreement. This reconfirms what we found in Panel A of Table \ref{stats_sentiment_user}. As for investors' experience: (i) that professional investors are more pessimistic than novice and intermediate investors; (ii) the novice and intermediate investors share the same pattern across time whereas professional investors have a substantially different pattern. For example, disagreement is increasing in the middle of January 2020 and stays up until the end of March. However, disagreement between novice and intermediate investors is pretty stable until the middle of February, where the market reaches a historical high, and then they have a sharp increase towards the middle of March, where the market had a record drop. As for investors' investment horizon: (i) position traders, long term traders, and swing traders share the same pattern for daily time series of average sentiment and disagreement, whereas the pattern of day trader investors is different. (ii) day trader investors are more pessimistic than other investors within this group. Consequently, disagreement is much higher, from the middle of January it is around $ 1 $ which is pretty high.

%Figure \ref{fig_sentiment_app_detailed_1} presents daily time series for average sentiment and disagreement within investors' philosophy. A visualization reveals that all investors, except growth, share the same opinion on sentiment and disagreement. This reconfirms what we found in Panel A of Table \ref{stats_sentiment_user}. 

%Although we have not reported here the same graph as Figure \ref{fig_sentiment_app_detailed_1} for investors' experience and investment horizon,\footnote{The results are available upon request.} howsoever, we observer the following messages. 

On February 19, 2020, the US stock market was at a historical high and followed by a record drop on March 23, 2020. We refer to these dates the good and bad state of the economy, respectively. By the end of March, the market has recovered some of its losses. We refer to this state, which is the last observation in our sample, as the "recovered" state of the economy. It is interesting to investigate the disagreement correlation matrix within investors with different investment philosophies, horizons, and experiences at different stages of the economy. The outcome of such an investigation potentially can help to identify the source of disagreement in these states of the economy. Consequently, it can provide good guidance for the designing ongoing economic policy response to the COVID-19 pandemic.

\begin{figure}[!ht]
	\centering
	\caption{Disagreement correlation matrix of users within their approach.}
	\caption*{\small{This figure has three panels to represent three different states of the economy during the COVID-19 pandemic: good, bad, and "recovered". Within each panel, there are two heatmaps of disagreement correlation matrix with (right) and without (left) rolling window. A user can specify his/her approach. There are five options for this section: fundamental, technical, growth, value, momentum. RW: rolling window.  }}
	\renewcommand{\thefigure}{\arabic{figure}}
	\begin{subfigure}[b]{1\textwidth}
		\centering
		\includegraphics[scale=0.3]{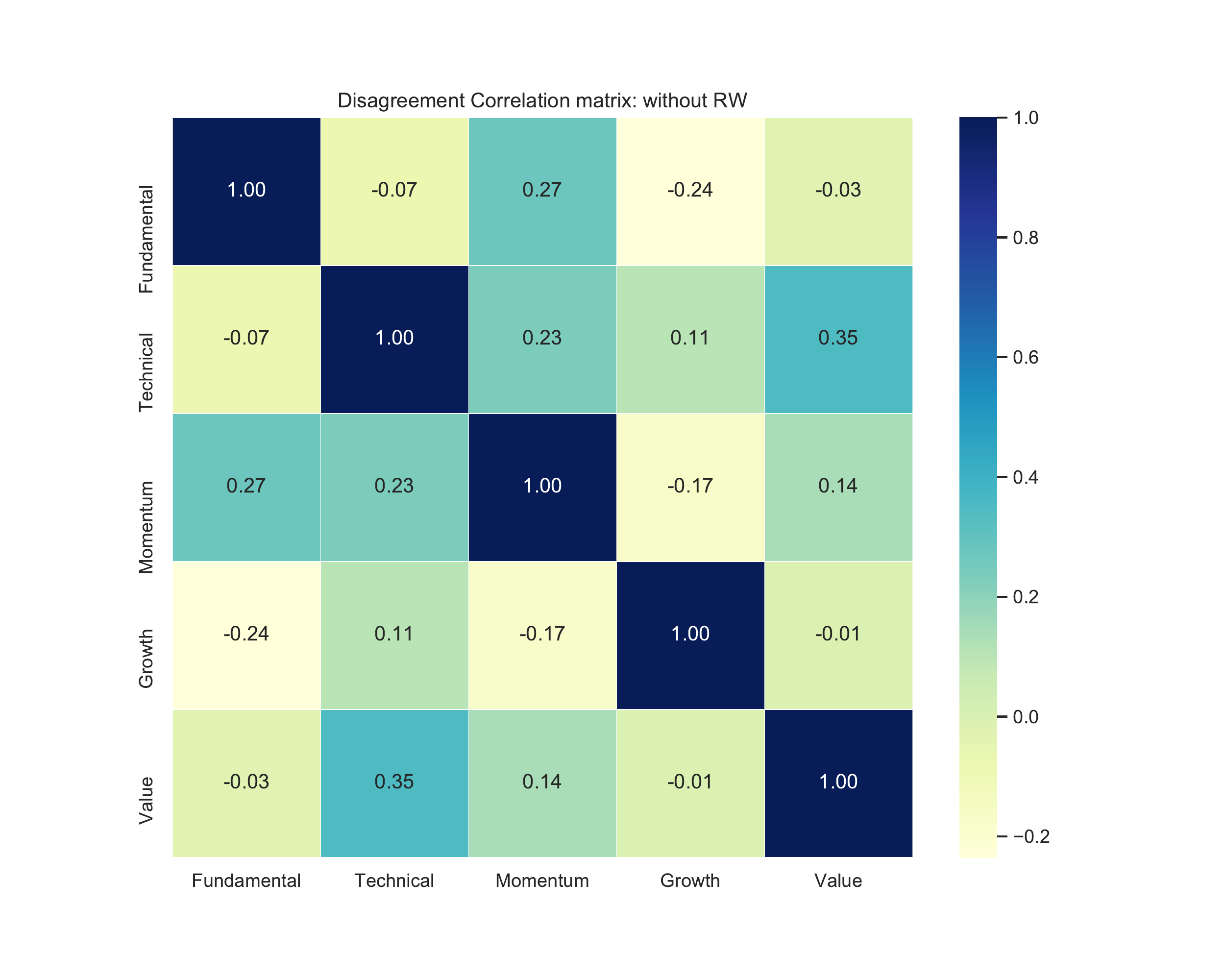}    
		\includegraphics[scale=0.3]{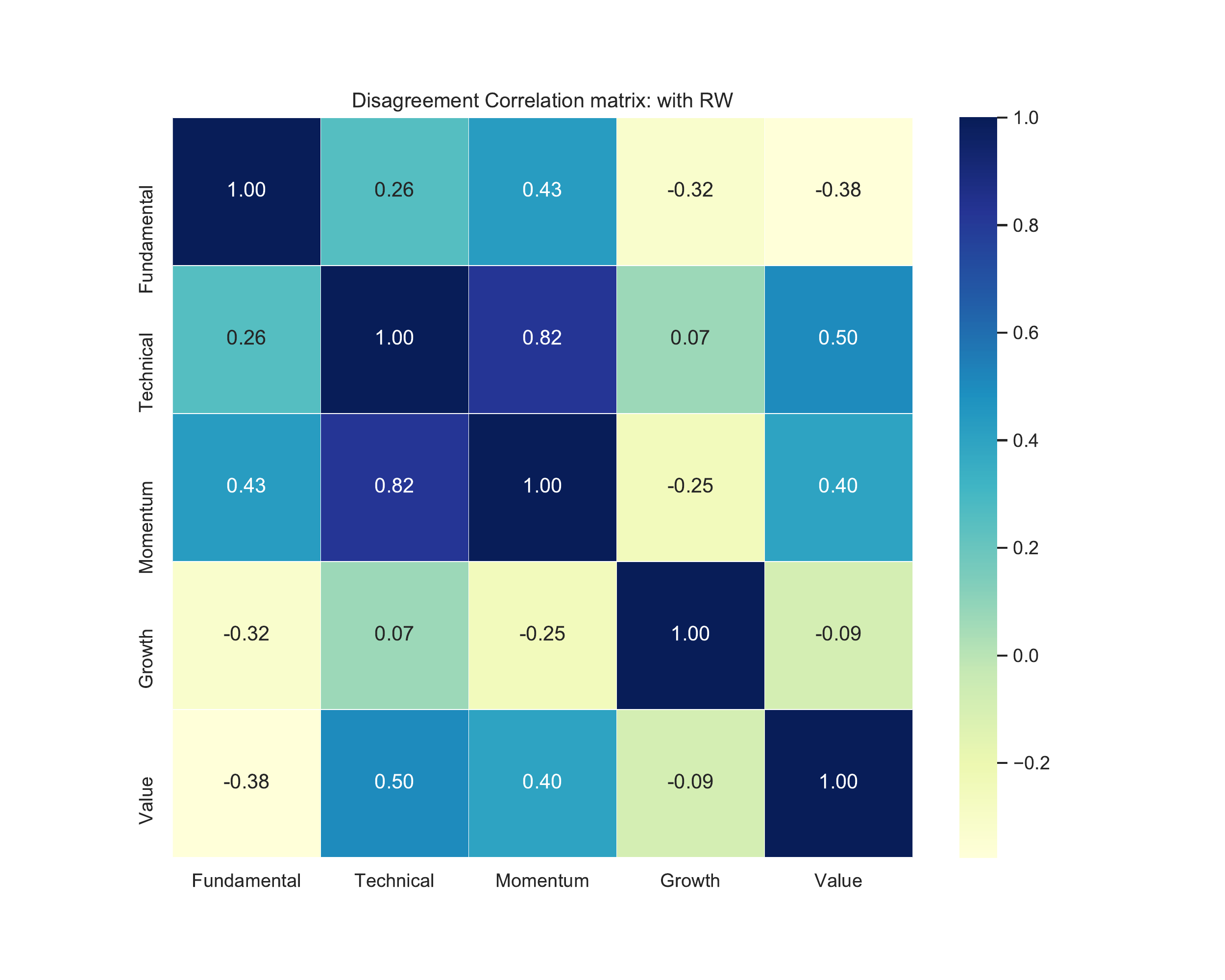}                               
		\caption{February 19, 2020: high market}
		%        \label{top10_repeared_wored}
	\end{subfigure} 
	
	\begin{subfigure}[b]{1\textwidth}
		\centering
		\includegraphics[scale=0.3]{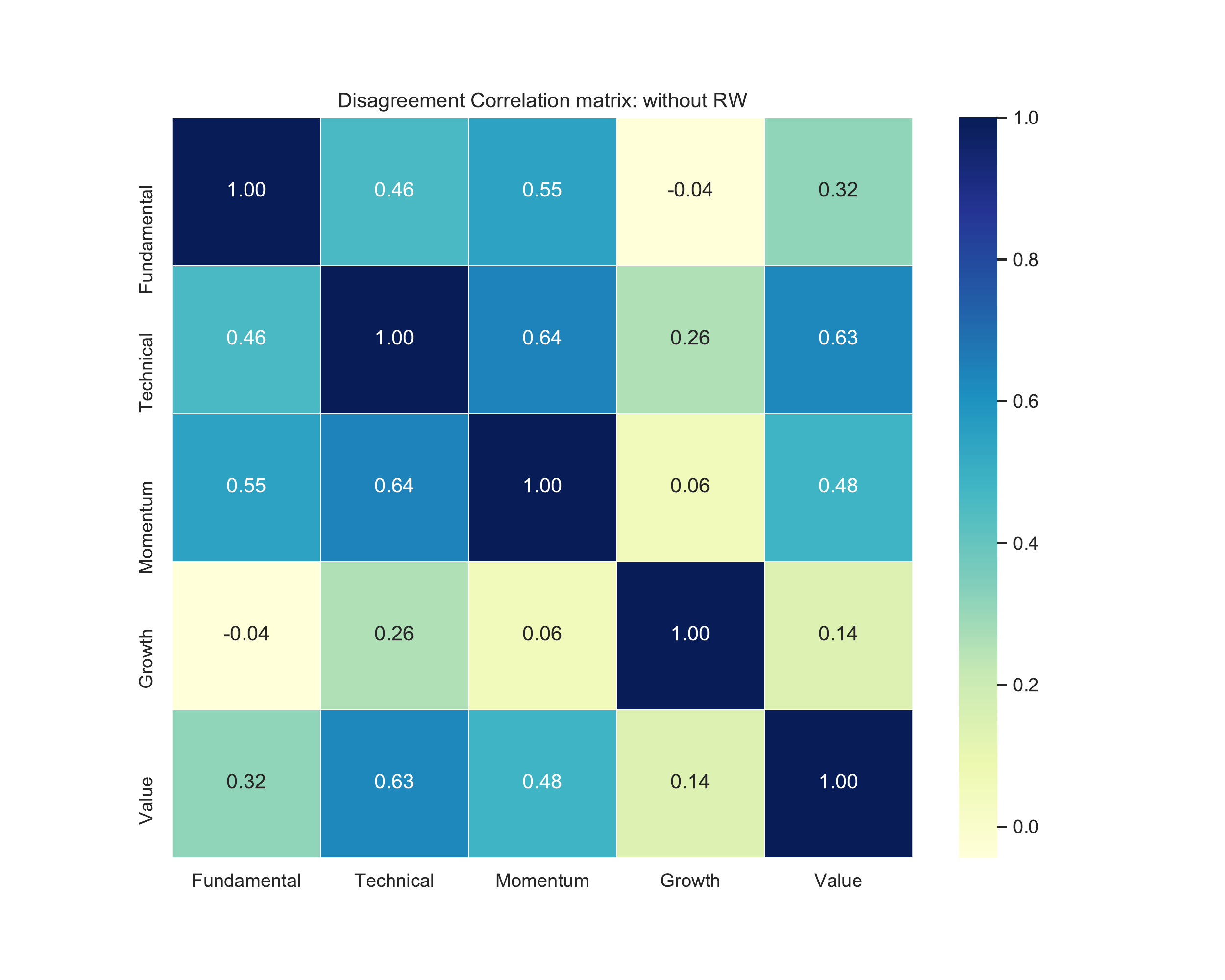}    
		\includegraphics[scale=0.3]{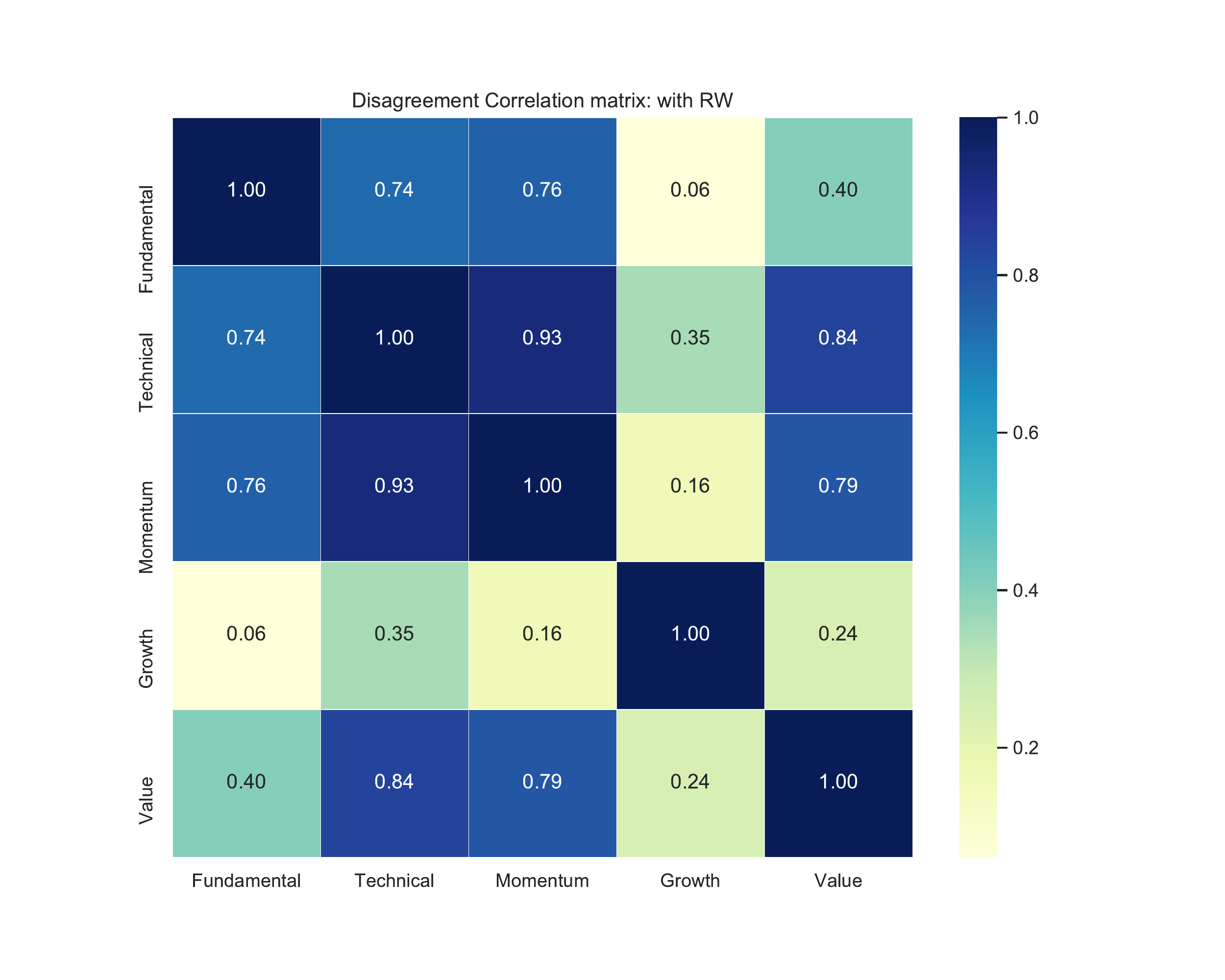}                               
		\caption{March 23, 2020: low market}
		%        \label{top10_repeared_wored}
	\end{subfigure}  
	
	\begin{subfigure}[b]{1\textwidth}
		\centering
		\includegraphics[scale=0.3]{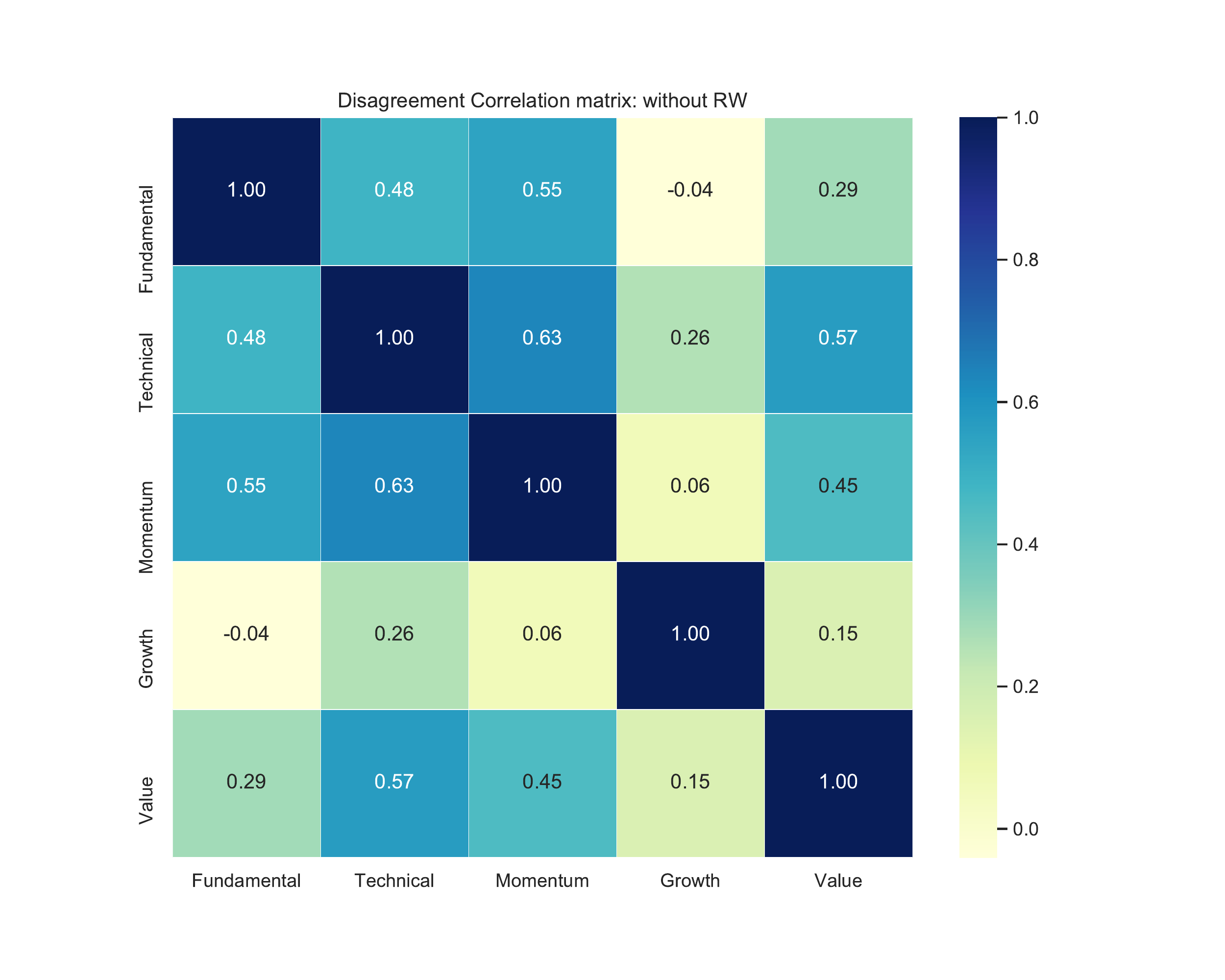}    
		\includegraphics[scale=0.3]{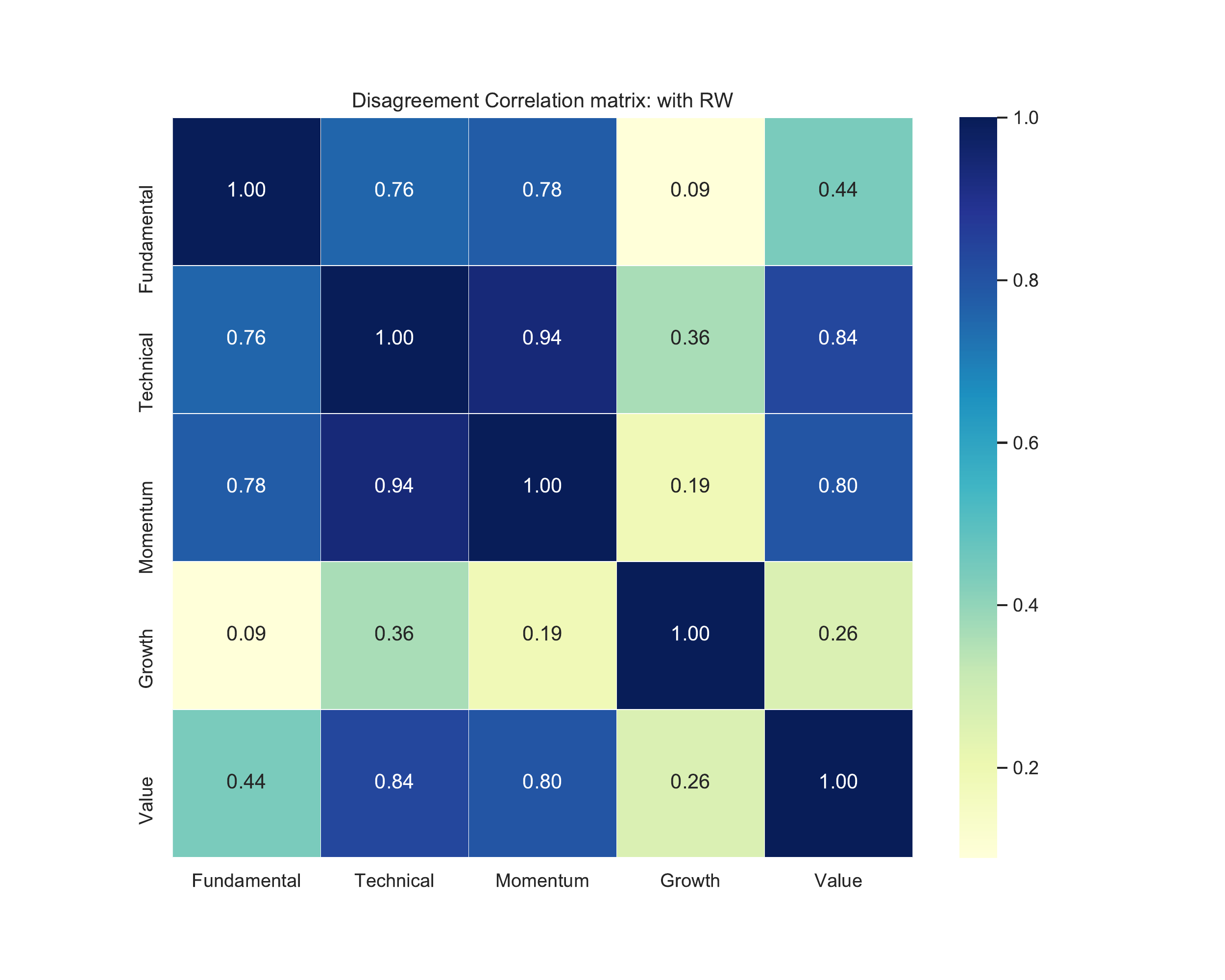}                               
		\caption{March 31, 2020: last day in sample}
		%        \label{top10_repeared_wored}
	\end{subfigure} 
	
	\label{fig_disagreement_App_Corr}
\end{figure}

Figure \ref{fig_disagreement_App_Corr} presents the disagreement correlation matrix across investors' philosophy at three states of the economy:  good (a), bad (b), and "recovered" (c). We use all observations in our sample from November 31, 2019, until February 19 (March 23), 2020 for calculating disagreement correlation matrix in the good (bad) state of the economy. In each panel, there are two heatmaps, with (right) and without (left) rolling window. A close inspection reveals three important findings. First, disagreement for growth investors in all three states of the economy is less correlated to investors with other investment philosophies. For example, in the good state of economy growth investors have a negative correlation with fundamental, momentum, and value investors. Second, in the good state of the economy disagreement correlation matrix has smaller values in comparison with the bad sate. For instance, the correlation between momentum and value investors increases from $ 0.40 $ to $ 0.79 $. Third, although the market gains back some of its losses by the end of March, however, the values of the disagreement correlation matrix are still high. This is consistent with what we find earlier, which is the disagreement decreases towards the end of the sample for all investment philosophies. The same results hold for investors' experience and investment horizon.\footnote{The results are available upon request.}

It has been documented that disagreement, volume, and volatility are highly correlated, especially during a bear market, see \cite{bollerslev2018volume}, among others. However, it is not clear what is the source of disagreement: information or interpretation of information (investment philosophy). \cite{cookson2018don} find that disagreement is evenly split between both sources of disagreement. Surprisingly, a conclusion from Figure \ref{fig_disagreement_App_Corr} is that the differential interpretation of information is not likely a source of disagreement during the COVID-19 pandemic. The main argument to support this claim is the high correlation among investment philosophies.

\subsection{Sectors}

\textit{Summary}. By exploiting the evolution of average sentiment and disagreement across and within sectors, in contrast to our finding across investors, we find that the pattern of sentiment and disagreement is heterogeneous across sectors. Moreover, the financial sector is the most pessimistic while healthcare is the most optimistic during the COVID-19 pandemic.\footnote{By pessimistic (optimistic) we mean investors post less bullish (bearish) messages and more bearish (bullish) messages.} The average sentiment for the financial sector is even entering within negative territory during March.

Table \ref{stats_day_sector_1} presents summary statistics on average sentiment broken down by sector.\footnote{Due to the low quality of posted messages about the Conglomerates sector, we exclude this sector form our analysis.} The Healthcare sector has the highest average number of messages per day, $ 3,910 $ messages, followed by financial and technology sectors with $ 4,125 $ and $ 1,997 $ messages, respectively. This is consistent with our original sample, except the financial sector had the highest number of posed messages, see Figure \ref{fig_top20_cashtages_sector}. Industrial goods and utilities sectors have the smallest average number of posted messages per day by $ 654 $ and $ 694 $ messages. 

Panel B and C of Table \ref{stats_day_sector_1} represent the descriptive statistics on average sentiment and disagreement per day per sector, respectively. A close inspection of the mean columns reveals two important discoveries. First, on average investors do not post the same number of bullish/bearish messages across sectors. Second, the financial sector is the most pessimistic, with $ 0.199 $ average sentiment per day, while healthcare is the most optimistic with $ 0.808 $ average sentiment per day during the COVID-19 pandemic. Consequently, the disagreement measure is the highest (lowest) for the financial (healthcare) sector where the average of disagreement is $ 0.938 (0.584) $ per day. After the financial sector, the consumer goods sector is the second most pessimistic sector where the average sentiment is $ 0.312 $. Investors tend to pose the same number of messages for industrial goods and services where the average sentiment per day is $ 0.480 $ and $ 0.408 $, respectively.

\begin{figure}[!ht]
	\centering
	\caption{Daily time series of sentiment and disagreement across sectors.}
	\caption*{\scriptsize{This figure exhibits the daily time series of sentiment and disagreement of users within their approach. A posted messages on StockTwits can include a cashtage, i.e., \$ticker, to specify the messages is related to which firm. By mentioning a ticker the posted message can belong to one of these sectors: basic materials, financial, consumer goods, healthcare, industrial goods, technology, utilities, service, and conglomerates.There are four options for this section: day trader, position trader, long term trader, and swing trader. We first label each bearish message as $ -1 $ and each bullish message as $ 1 $. We then take the arithmetic average of these classifications at the $ group1\times day\times group2 $ level: $ \text{AvgSentiment}_{itg}= \dfrac{N^{Bullish}_{itg} - N^{bearish}_{itg}}{N^{Bullish}_{itg} + N^{bearish}_{itg}}$. Group1 can be either all firms, sectors, industries, or specific firm, sector, or industry.	Group2 can either be all investors or investors with a given investment philosophy, experience, or holding period (investment horizon) level. Furthermore, to follow the sentiment's trend, we use rolling windows of past seven days $ \text{AvgSentiment}_{itg} $ for measuring the sentiment at each day. Disagreement is calculated by: $\text{Disagreement}_{itg}= \sqrt{1-\text{AvgSentiment}_{itg}^2} $.}}
	\renewcommand{\thefigure}{\arabic{figure}}
	\begin{subfigure}[b]{1\textwidth}
		\centering
		\includegraphics[scale=0.15]{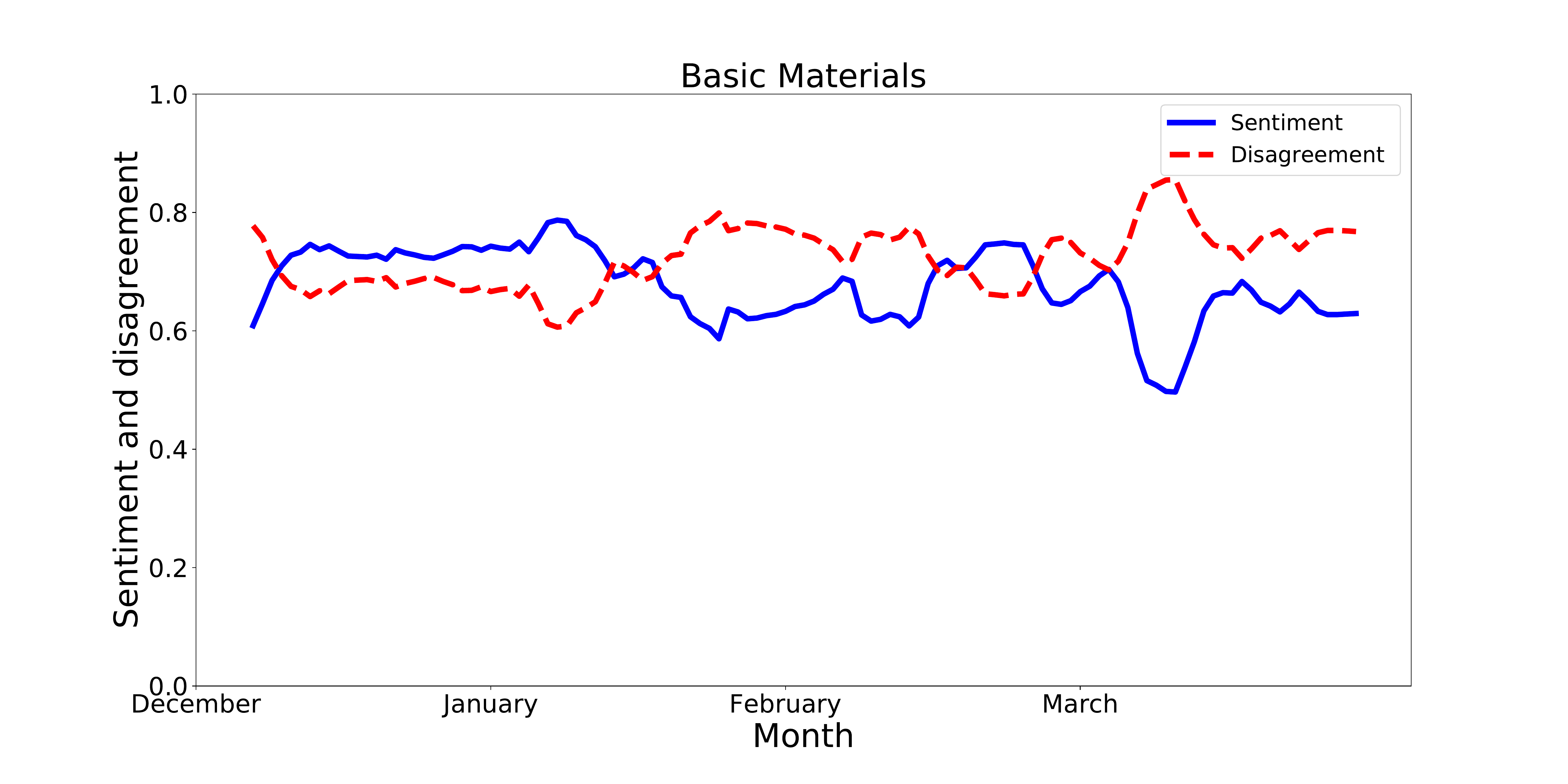}   
		\includegraphics[scale=0.15]{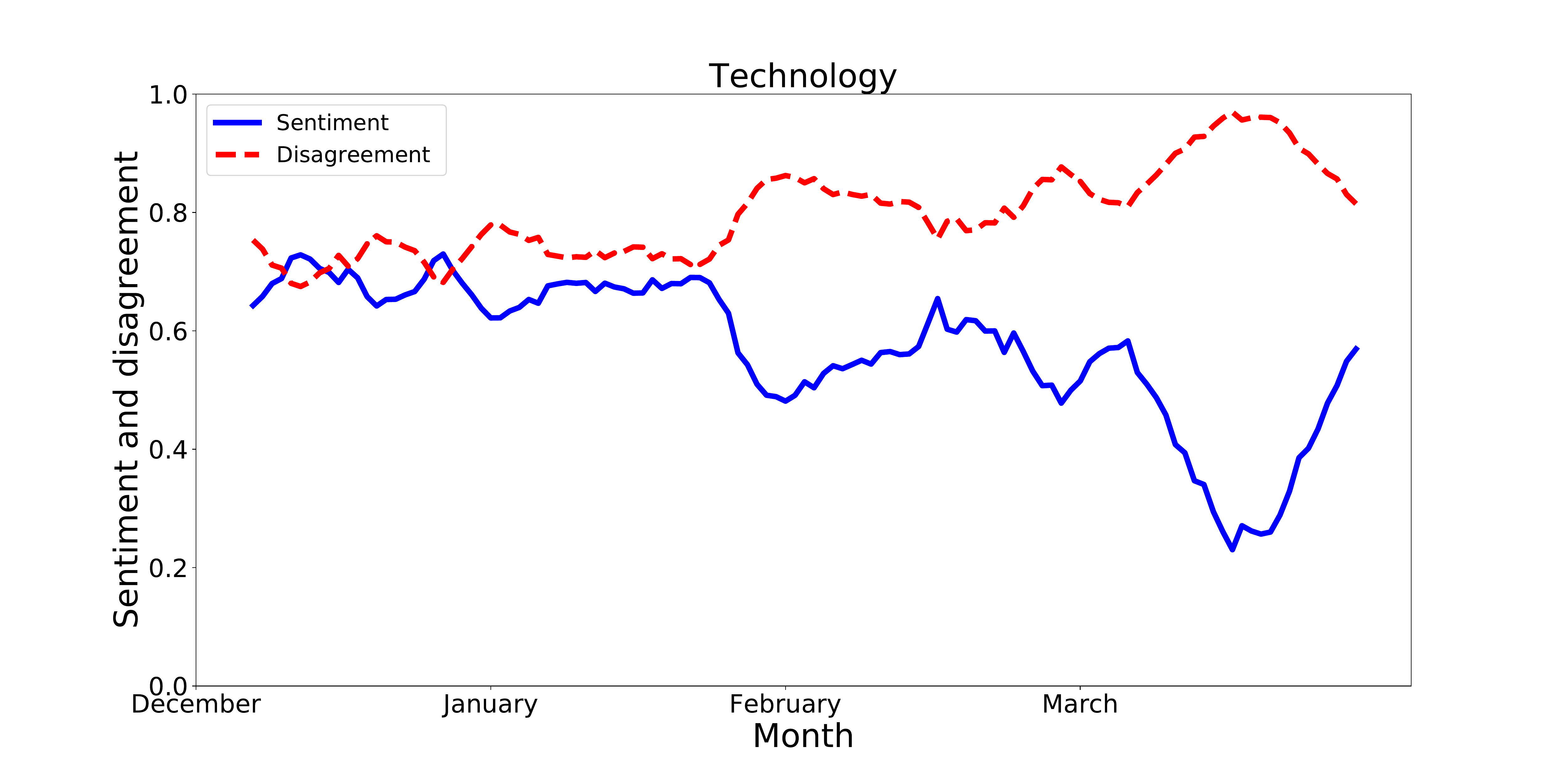}                             
		%             \caption{Day-of-week frequency distribution of messages posted to StockTwits}
		%        \label{top10_repeared_wored}
	\end{subfigure} 	
	\begin{subfigure}[b]{1\textwidth}
		\centering
		\includegraphics[scale=0.15]{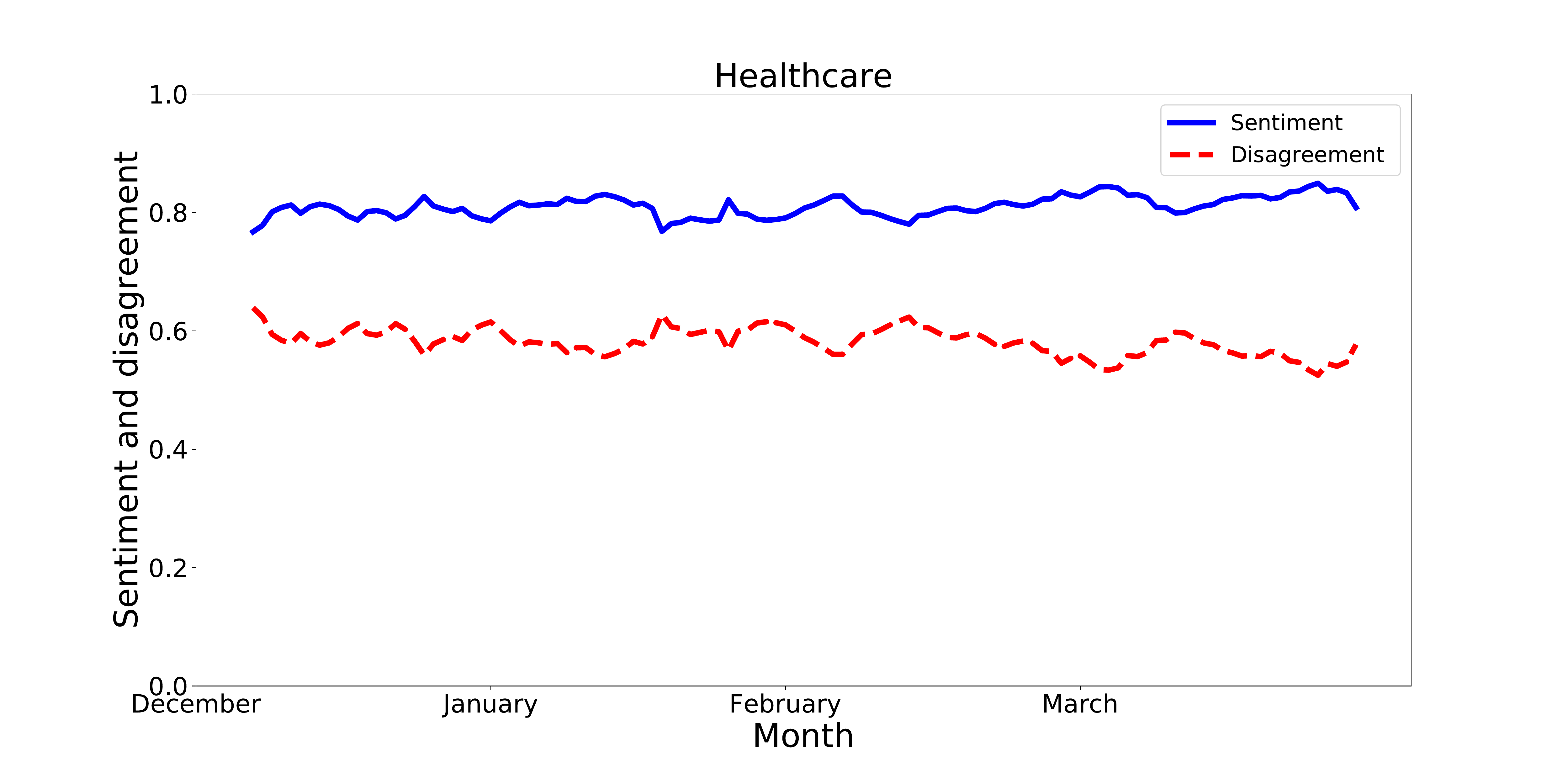}    
		\includegraphics[scale=0.15]{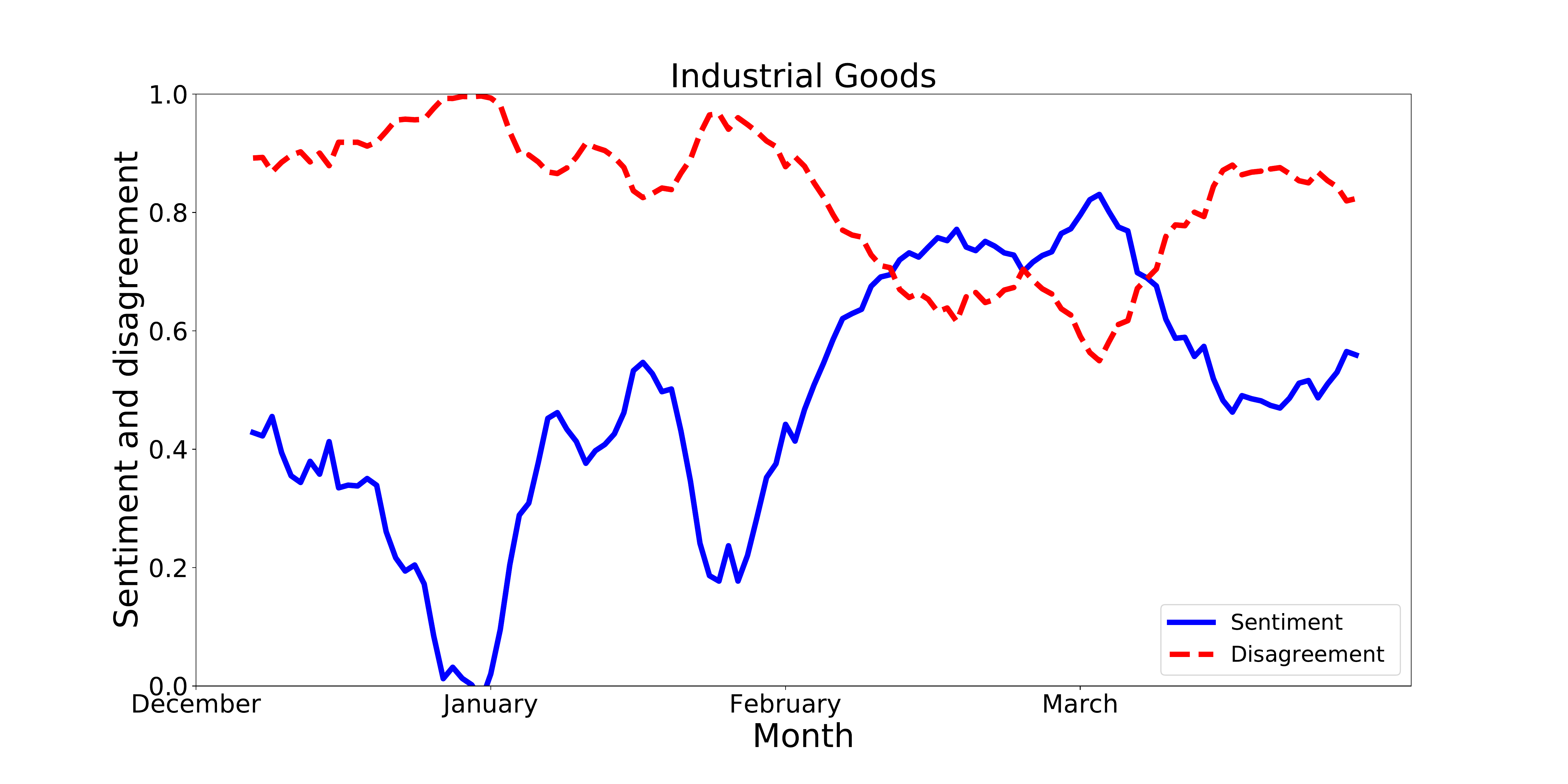}                              
		%		\caption{bla bla bla ...}
		%		\label{top10_repeared_wored_percentage}
	\end{subfigure} 
	
	\begin{subfigure}[b]{1\textwidth}
		\centering
		\includegraphics[scale=0.15]{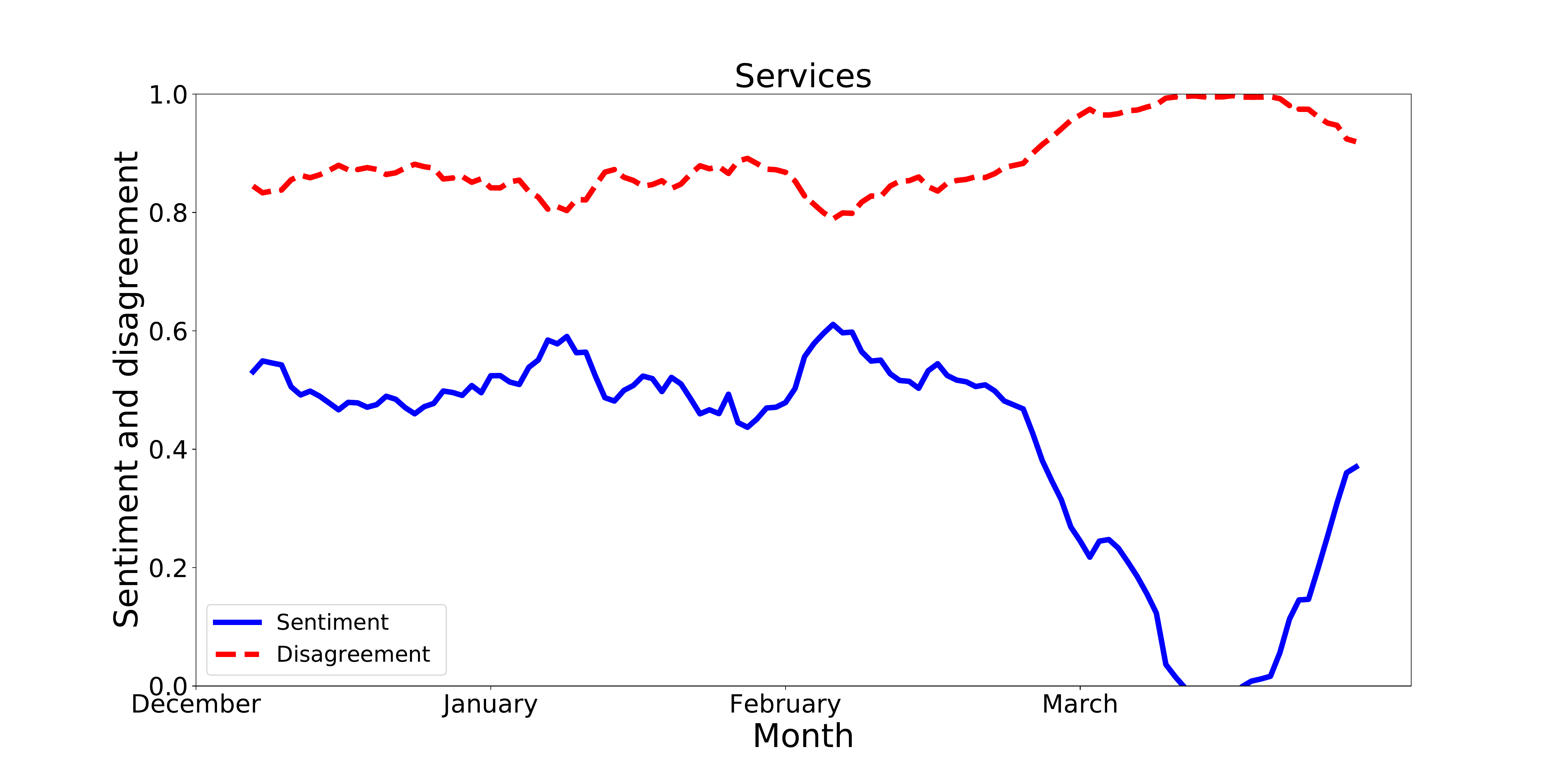}     
		\includegraphics[scale=0.15]{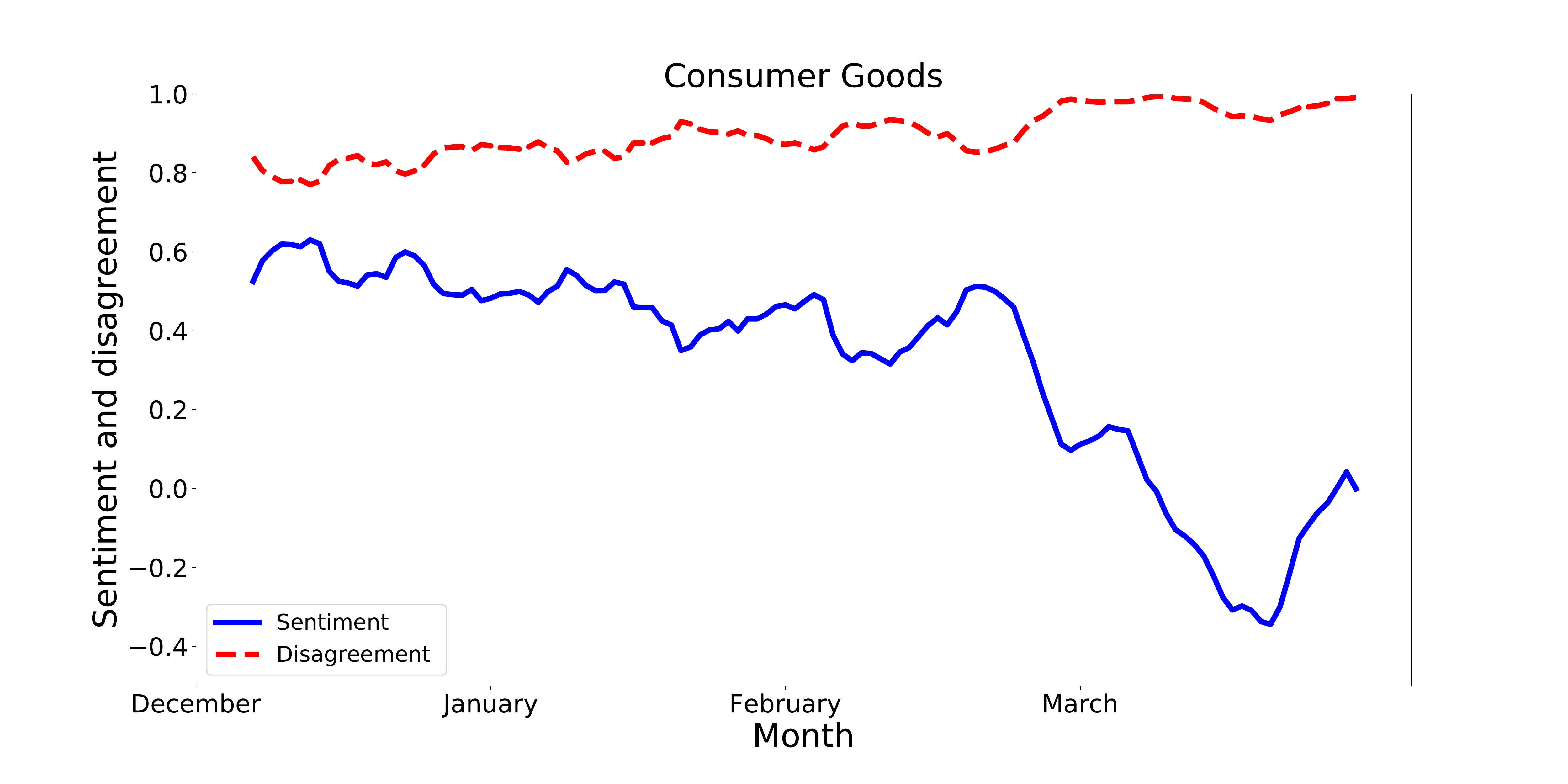}                            
		%		\caption{bla bla bla ...}
		%		\label{top10_repeared_wored_percentage}
	\end{subfigure} 
	\begin{subfigure}[b]{1\textwidth}
		\centering
		\includegraphics[scale=0.15]{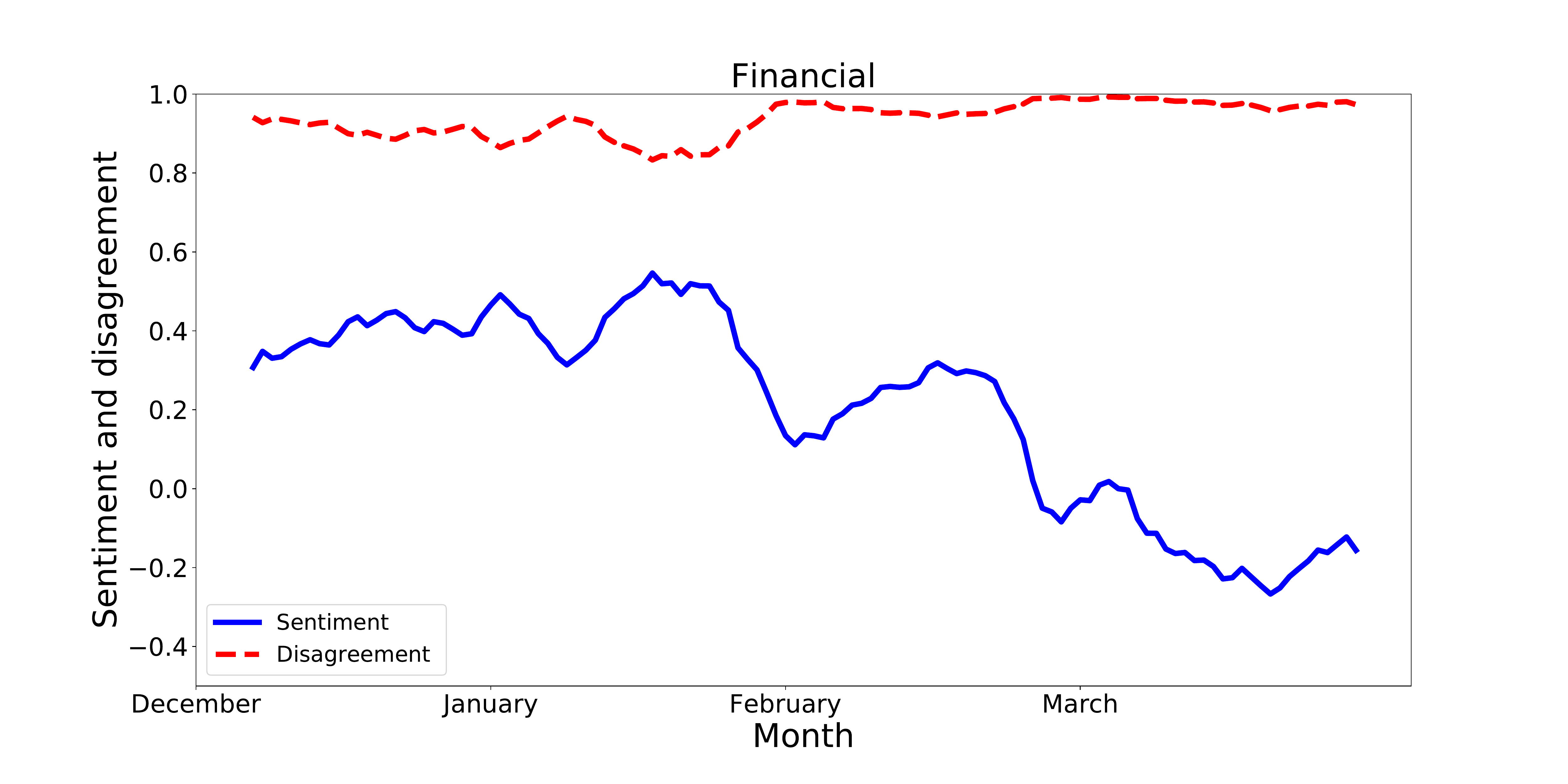}    
		\includegraphics[scale=0.15]{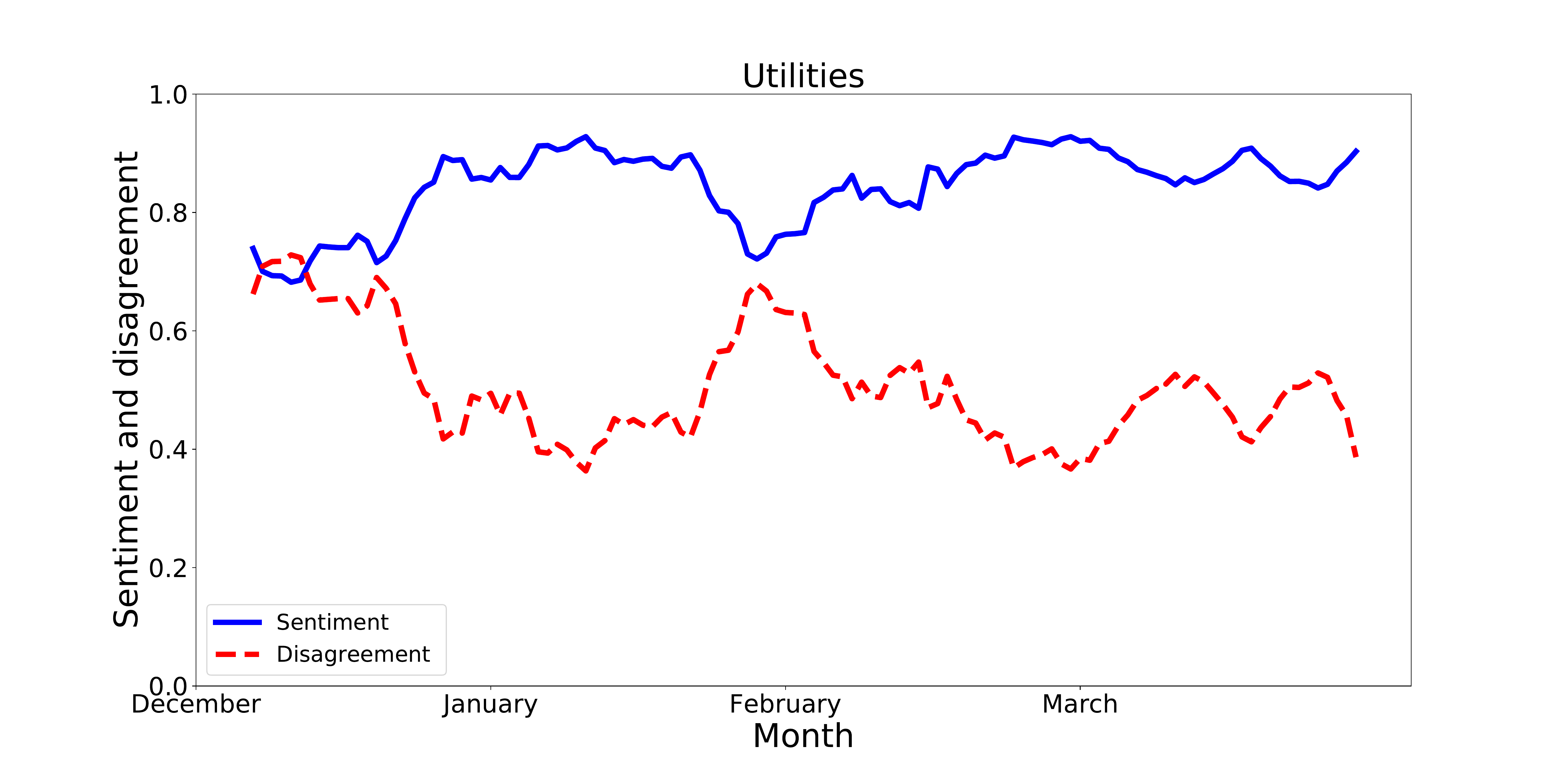}                                
		%             \caption{Day-of-week frequency distribution of messages posted to StockTwits}
		%        \label{top10_repeared_wored}
	\end{subfigure} 
	\label{fig_sentiment_sector_1}
\end{figure}

Figure \ref{fig_sentiment_sector_1} exhibits daily time series of sentiment and disagreement across sectors. A visualization re-confirm our findings from Table \ref{stats_day_sector_1}. Moreover, this figure allows us to see their evolution which is not possible from Table  \ref{stats_day_sector_1}. Time series of sentiment and disagreement differently evolves across sectors. Specifically, we can see that the time series of sentiment for financial, consumer goods, technology, and services sectors have downward trends while utilities, industrial goods sectors have time series with an upward trend. The trend of sentiment for the healthcare and basic material is almost flat where the sentiment and disagreement for the basic material sector is more volatile than the rest of the sectors. 

%Another interesting finding from Figure \ref{fig_sentiment_sector_1} is related to the evolution of industrial sector's sentiment/disagreement. The sentiment drops much earlier than other sectors, beginning of December. 

In a good economy state, each asset class or industry moves independently, i.e., there are small price correlations among different classes of assets or sectors. To explore this behavior for sentiment and disagreement, we calculate the disagreement correlation matrix across all sectors on three economy states: good, bad, and recovered. 

Panel A of Table \ref{corr_sector_} presents the disagreement correlation matrix in the good state of the economy across sectors. Among others, we can see that disagreement between finance and technology industries is high and positive, $0.667 $, while the correlation of disagreement between finance and healthcare sectors is small and negative, $ -0.091 $. Disagreement in the technology sector is negatively correlated with industrial goods and services sectors. Moreover, disagreement of healthcare is positively correlated with its counterparts in all other sectors, except for the financial sector. 

Panel B of Table \ref{corr_sector_} reveals the disagreement correlation matrix in the bad state of the economy across sectors. In this state of the economy, we can see that in contrast to the good state of the economy, the correlation of the healthcare sector is negative with most of the other sectors. The disagreement correlation matrix is pretty stable by going toward end of the sample, see Panel C.

%
%\subsection{The source of disagreement}
%
%\hasan{During those times, agents trade mostly by looking at each other’s strategies rather than information}
%

\section{Conclusion}

To explore the investor beliefs, sentiment and disagreement, on stock market returns during the COVID-19 pandemic, we conducted a comprehensive analysis of all posted messages on the StockTwits platform between November 30, 2020, and March 31, 2020. We established several empirical facts about sentiment and disagreement of investors that post messages on the StockTwits platform. First, the daily time series of sentiment and disagreement is not a stationary process. Second, there is a $ V-$shape ($ \Lambda-$shape ) in sentiment (disagreement) between February 19, 2020 and March 31, 2020. Third, the pattern of sentiment and disagreement is homogeneous for investors across and within investment philosophy, horizons, and experiences. Fourth, the pattern of sentiment and disagreement is heterogeneous across sectors. Fifth, the financial sector is the most pessimistic while healthcare is the most optimistic during the COVID-19 pandemic.

There are several unexplored and exciting directions for further investigation on the StockTwits dataset. For example, John Maynard Keynes claims that when there is too much activity in the market, e.g., during the crisis, investors/agents trade mostly by looking at each other’s strategies rather than information. Therefore, it would be interesting to explore what kind of investor influences other investors and how he/she dominates the market. To do so, one can use a multivariate time-dependent point process such as Hawkes Processes, see \cite{ait2015modeling}, among others.

\begin{landscape}
\begin{table}[H]
	\centering
		\ra{0.7}
	%   \fontsize{30pt}{30pt}
	%    \selectfont
	%	\large
	\caption{Characteristics of StockTwits Data }
	\caption*{\textbf{Note:} 
		In this table, we report summary statistics for the StockTwits data. Panel A presents summary information on coverage by stock and user, as well as user-level information. Panel B presents frequency distributions of messages' basic features: number of words, characters, Stopdwrods, and Cashtages, as well as, the average length of words.}
	\begin{adjustbox}{max width= 30cm}
		\begin{tabular}{lcccccccr}
			%			{Panel A: Sentiment stats}&\\
			\toprule
			\multicolumn{8}{l}{Panel A: Characteristics of users and messages}\\
			%			\midrule
			Approach   & count & mean  & std   & min   & 25\%  & 50\%  & 75\%  & max \\
			\midrule
			Number of messages per stock & 10,715 & 1,011.77 & 12,729 & 1     & 6     & 63    & 272   & 1,087,183 \\
			Number of messages per user & 179,468 & 6.66  & 21.42 & 1     & 1     & 2     & 6     & 5000 \\
			Number of messages per stock per day & 451,965 & 23.99 & 258.65 & 1     & 1     & 3     & 8     & 26,424 \\
			Number of followers user has & 179,468 & 121.37 & 2,825.24 & -1    & 0     & 1     & 5     & 308,789 \\
			Number of people user follows & 179,468 & 23.92 & 103.78 & -3    & 0     & 3     & 24    & 10,000 \\
			Number of ideas user has & 179,468 & 628.7 & 7,136.09 & 0     & 2     & 37    & 279   & 2,126,704 \\
			Number of likes user has & 179,468 & 487.07 & 3,096.58 & 0     & 1     & 22    & 186   & 809,351 \\
			%			\bottomrule
			%			{Panel B}\\
			\midrule
			\multicolumn{8}{l}{Panel B: Basic feature of StockTwits's messages}\\
			Holding Period   & count & mean  & std   & min   & 25\%  & 50\%  & 75\%  & max \\
			\midrule
			Number of words per message & 3,676,169 & 15.55 & 18.85 & 1     & 5     & 10    & 19    & 605 \\
			Number of characters per message & 3,676,169 & 88.05 & 109.77 & 2     & 29    & 55    & 106   & 3,617 \\
			Average word's length per message & 3,676,169 & 5.18  & 5.11  & 1.01  & 4     & 4.5   & 5     & 1,050 \\
			Number of Stopdwrods per message & 3,676,169 & 4.91  & 7.4   & 0     & 1     & 3     & 6     & 110 \\
			Number of Cashtages per message & 3,676,169 & 1.15  & 0.52  & 0     & 1     & 1     & 1     & 250 \\
			\bottomrule
		\end{tabular}%
	\end{adjustbox}
	\label{tab_characteristics_message_users}%
\end{table}%
\end{landscape}

% Table generated by Excel2LaTeX from sheet 'stats_messages_Table1_PanelB_al'
\begin{table}[H]
	\centering
	\ra{0.7}
	%   \fontsize{30pt}{30pt}
	%    \selectfont
	\large
	\caption{Frequencies of User Profile Characteristics }
	\caption*{\textbf{Note:} This table presents frequency distributions of users and messages posted by investment philosophy, holding period, and experience, which are observed user profile characteristics.}
	\begin{adjustbox}{max width=\textwidth}
		\begin{tabular}{lcccr}
			\toprule
			\multicolumn{5}{l}{Panel A: investors' approach }\\
			Approach & Num. Users & Percent Users & Num. Messages & Percent Messages \\
			\midrule
    	Technical & 10,615 & 5.92\% & 356,522 & 9.70\% \\
    	Growth & 6,175  & 3.44\% & 251,412 & 6.84\% \\
    	Momentum & 6,126  & 3.42\% & 240,505 & 6.54\% \\
    	Fundamental & 4,459  & 2.48\% & 131,304 & 3.57\% \\
    	Value & 3,347  & 1.86\% & 95,407 & 2.60\% \\
    	Global Macro & 1,394  & 0.78\% & 20,655 & 0.56\% \\
    	Not Classified & 147,352 & 82.10\% & 2,580,364 & 70.19\% \\
    		\rowcolor[gray]{0.9}Total & 179,468 & 100.00\% & 3,676,169 & 100.00\% \\
			\toprule
			\multicolumn{5}{l}{Panel B: investors' horizon}\\
			Holding Period & Num. Users & Percent Users & Num. Messages & Percent Messages \\
			\midrule
			Swing Trader & 12,291 & 6.85\% & 433,131 & 11.78\% \\
			Long Term Investor & 8,099  & 4.51\% & 198,742 & 5.41\% \\
			Day Trader & 6,479  & 3.61\% & 247,245 & 6.73\% \\
			Position Trader & 5,351  & 2.98\% & 182,010 & 4.95\% \\
			Not Classified & 147,248 & 82.05\% & 2,615,041 & 71.13\% \\
				\rowcolor[gray]{0.9}Total & 179,468 & 100.00\% & 3,676,169 & 100.00\% \\
			\toprule
			\multicolumn{5}{l}{Panel C: investors' experience}\\
			Experience & Num. Users & Percent Users & Num. Messages & Percent Messages \\
			\midrule
			Novice & 8,829  & 4.92\% & 187,436 & 5.10\% \\
			Intermediate & 15,177 & 8.46\% & 580,374 & 15.79\% \\
			Professional & 8,468  & 4.72\% & 320,820 & 8.73\% \\
			Not Classified & 146994 & 81.91\% & 2,587,539 & 70.39\% \\
			\rowcolor[gray]{0.9}Total & 179,468 & 100\% & 3,676,169 & 100.00\% \\
			\bottomrule
		\end{tabular}%
	\end{adjustbox}
	\label{table:user_message_approach}%
\end{table}%

%\begin{landscape}
\begin{table}[H]
	\centering
		\ra{0.9}
	%   \fontsize{30pt}{30pt}
	%    \selectfont
	%	\large
	\caption{Sentiment profile of users }
	\caption*{\textbf{Note:} In this table, we present summary statistics for our sentiment across all investors. Specifically, this table presents summary information on the StockTwits measure of sentiment across groups with different investment philosophies (Panel A ), investment horizon (Panel B), and experience (Panel C).}
	\begin{adjustbox}{max width= 30cm}
		\begin{tabular}{lcccccccr}
			%			{Panel A: Sentiment stats}&\\
			\toprule
			\multicolumn{8}{l}{Panel A: Descriptive statistics of sentiment (Approach)}\\
			%			\midrule
			Approach   & count & mean  & std   & min   & 25\%  & 50\%  & 75\%  & max \\
			\midrule
			Fundamental & 30,067 & 0.653 & 0.701 & -1    & 1     & 1     & 1     & 1 \\
			Technical & 64,977 & 0.612 & 0.72  & -1    & 0.647 & 1     & 1     & 1 \\
			Momentum & 34,584 & 0.657 & 0.677 & -1    & 0.818 & 1     & 1     & 1 \\
			Growth & 30,067 & 0.653 & 0.701 & -1    & 1     & 1     & 1     & 1 \\
			\rowcolor[gray]{0.9} Value & 24,070 & 0.721 & 0.641 & -1    & 1     & 1     & 1     & 1 \\
			%			\bottomrule
			%			{Panel B}\\
			\midrule
			\multicolumn{8}{l}{Panel B: Descriptive statistics of sentiment (Holding Period)}\\
			Holding Period   & count & mean  & std   & min   & 25\%  & 50\%  & 75\%  & max \\
			\midrule
			Swing Trader & 60,818 & 0.672 & 0.662 & -1    & 0.889 & 1     & 1     & 1 \\
			Long Term Investor & 41,809 & 0.685 & 0.68  & -1    & 1     & 1     & 1     & 1 \\
			\rowcolor[gray]{0.9} Day Trader & 50,471 & 0.372 & 0.873 & -1    & -1    & 1     & 1     & 1 \\
			Position Trader & 42,037 & 0.668 & 0.692 & -1    & 1     & 1     & 1     & 1 \\
			\midrule
			\multicolumn{8}{l}{Panel C: Descriptive statistics of sentiment (Experience)}\\
			Experience   & count & mean  & std   & min   & 25\%  & 50\%  & 75\%  & max \\
			\midrule
			 Intermediate & 67,930 & 0.714 & 0.616 & -1    & 0.956 & 1     & 1     & 1 \\
			Novice & 33,814 & 0.738 & 0.611 & -1    & 1     & 1     & 1     & 1 \\
			\rowcolor[gray]{0.9} Professional & 75,222 & 0.427 & 0.846 & -1    & -0.333 & 1     & 1     & 1 \\
			\bottomrule
		\end{tabular}%
	\end{adjustbox}
	\label{stats_sentiment_user}%
\end{table}%
%\end{landscape}

%\begin{landscape}
\begin{table}[H]
	\centering
		\ra{0.7}
	%   \fontsize{30pt}{30pt}
	%    \selectfont
	%	\large
	\caption{Sectors' Characteristics: summary statistics per day per sector }
	\caption*{\textbf{Note:} In this table, we present summary statistics for sentiment (Panel B), disagreement (Panel C), the number of messages (Panel A). Count column represents number of days in sample. }
	\begin{adjustbox}{max width= \textwidth}
		\begin{tabular}{lcccccccr}
			%			{Panel A: Sentiment stats}&\\
			\toprule
			\multicolumn{8}{l}{Panel A: Descriptive statistics of count}\\
			Experience   & count & mean  & std   & min   & 25\%  & 50\%  & 75\%  & max \\
			\midrule
			Basic Materials & 123   & 949.69 & 704.514 & 18    & 220.5 & 997.5 & 1363  & 3,849 \\
			Technology & 123   & 1,997.23 & 1,349.075 & 51    & 463.25 & 2,249  & 2,872.5 & 5,457 \\
			\rowcolor[gray]{0.9}Healthcare & 123   & 3,910.651 & 3,733.966 & 81    & 685   & 3,511.5 & 4,557.5 & 17,582 \\
			Industrial Goods & 123   & 654.183 & 640.633 & 5     & 159.5 & 464.5 & 914.5 & 3,708 \\
			Services & 123   & 1,466.905 & 1,084.915 & 45    & 313.75 & 1,646  & 2,097  & 5,875 \\
			Consumer Goods & 123   & 1,649.825 & 1,704.921 & 30    & 381.25 & 1,329  & 2,373.5 & 10,816 \\
			\rowcolor[gray]{0.9} Financial & 123   & 4,125  & 3,274.753 & 99    & 1,910  & 3,253  & 5,184  & 14,286 \\
			Utilities & 123   & 694.524 & 794.683 & 14    & 171   & 415   & 884   & 4,033 \\
			\midrule
			\multicolumn{8}{l}{Panel B: Descriptive statistics of sentiment}\\
			%			\midrule
			Approach   & count & mean  & std   & min   & 25\%  & 50\%  & 75\%  & max \\
			\midrule
			Basic Materials & 123   & 0.669 & 0.116 & 0.222 & 0.605 & 0.69  & 0.741 & 0.879 \\
			Technology & 123   & 0.571 & 0.159 & 0.167 & 0.486 & 0.608 & 0.689 & 0.85 \\
			\rowcolor[gray]{0.9} Healthcare & 123   & 0.808 & 0.048 & 0.554 & 0.782 & 0.814 & 0.838 & 0.902 \\
			Industrial Goods & 123   & 0.480  & 0.261 & -0.111 & 0.31  & 0.491 & 0.691 & 0.91 \\
			Services & 123   & 0.408 & 0.206 & -0.185 & 0.288 & 0.467 & 0.551 & 0.729 \\
			Consumer Goods & 123   & 0.312 & 0.294 & -0.452 & 0.154 & 0.39  & 0.532 & 0.72 \\
			\rowcolor[gray]{0.9} Financial & 123   & 0.199 & 0.279 & -0.438 & -0.032 & 0.257 & 0.426 & 0.654 \\
			Utilities & 123   & 0.844 & 0.100   & 0.445 & 0.784 & 0.871 & 0.921 & 1 \\
			%			\bottomrule
			%			{Panel B}\\
			\midrule
			\multicolumn{8}{l}{Panel C: Descriptive statistics of disagreement}\\
			Holding Period   & count & mean  & std   & min   & 25\%  & 50\%  & 75\%  & max \\
			\midrule
			Basic Materials & 123   & 0.728 & 0.095 & 0.477 & 0.672 & 0.724 & 0.797 & 0.975 \\
			Technology & 123   & 0.800   & 0.101 & 0.527 & 0.725 & 0.794 & 0.874 & 0.986 \\
			\rowcolor[gray]{0.9} Healthcare & 123   & 0.584 & 0.062 & 0.431 & 0.546 & 0.581 & 0.623 & 0.833 \\
			Industrial Goods & 123   & 0.825 & 0.146 & 0.414 & 0.722 & 0.871 & 0.951 & 1 \\
			Services & 123   & 0.886 & 0.075 & 0.684 & 0.835 & 0.884 & 0.958 & 1 \\
			Consumer Goods & 123   & 0.900   & 0.080  & 0.694 & 0.846 & 0.914 & 0.975 & 1 \\
			\rowcolor[gray]{0.9} Financial & 123   & 0.938 & 0.062 & 0.757 & 0.901 & 0.959 & 0.989 & 1 \\
			Utilities & 123   & 0.505 & 0.152 & 0.000     & 0.389 & 0.492 & 0.62  & 0.895 \\
			\bottomrule
		\end{tabular}%
	\end{adjustbox}
	\label{stats_day_sector_1}%
\end{table}%
%\end{landscape}

\begin{landscape}
\begin{table}[H]
	\centering
	\ra{0.5}
	%   \fontsize{30pt}{30pt}
	%    \selectfont
	%	\large
	\caption{Disagreement correlation matrix of sectors on different sates of the economy. }
	\caption*{\textbf{Note:} This table presents the disagreement correlation matrix on different states of the economy: (Panel A) February 19, 2020, (Panel B) March 23, 2020, and March 31, 2020, during the COVID-19 pandemic. There are eight different sectors. BM: Basic Materials. T: Technology. H: Healthcare. IG: Industrial Goods. S: Services. CG: Consumer Goods. F: Financial. U: Utilities. RW: rolling window. }
	\begin{adjustbox}{max width= 1.23\textwidth}
		\begin{tabular}{lcccccccr}
			%			{Panel A: Sentiment stats}&\\
			\toprule
			\multicolumn{8}{l}{Panel A: Correlation matrix on February 19, 2020: high market}\\
			   &Basic Materials	&Technology	&Healthcare	&Industrial Goods&	Services	&Consumer Goods &	Financial&	Utilities \\
			\midrule
  		  Basic Materials & 1     &       &       &       &       &       &       &  \\
  		 Technology & 0.603 & 1     &       &       &       &       &       &  \\
  		 Healthcare & 0.485 & 0.147 & 1     &       &       &       &       &  \\
  		 Industrial Goods & $ -0.255 $ & -0.444 & 0.012 & 1     &       &       &       &  \\
  		 Services & 0.081 &$  -0.242 $ & 0.303 & 0.452 & 1     &       &       &  \\
  		 Consumer Goods & 0.641 & 0.584 & 0.222 & $ -0.478 $ & $ -0.229 $ & 1     &       &  \\
  		 Financial & 0.182 & 0.667 & $ -0.091 $ & $ -0.509 $ & $ -0.38 $ & 0.124 & 1     &  \\
  		 Utilities & 0.210 & 0.176 & 0.327 & 0.108 & 0.379 & $ -0.382 $ & 0.312 & 1 \\
			\midrule
			\multicolumn{8}{l}{Panel A: Correlation matrix on low market}\\
			&Basic Materials	&Technology	&Healthcare	&Industrial Goods&	Services	&Consumer Goods &	Financial&	Utilities \\
			\midrule
			Basic Materials & 1     &       &       &       &       &       &       &  \\
			Technology & 0.561 & 1     &       &       &       &       &       &  \\
			Healthcare & 0.154 &$  -0.264 $ & 1     &       &       &       &       &  \\
			Industrial Goods & $ -0.173 $ & $ -0.388 $ & 0.44  & 1     &       &       &       &  \\
			Services & 0.429 & 0.611 & $ -0.449 $ & $ -0.282 $ & 1     &       &       &  \\
			Consumer Goods & 0.651 & 0.703 & $ -0.359 $ & $ -0.584 $ & 0.635 & 1     &       &  \\
			Financial & 0.285 & 0.696 & $ -0.438 $ & $ -0.668 $ & 0.422 & 0.515 & 1     &  \\
			Utilities & 0.113 & $ -0.177 $ & 0.487 & 0.391 &$  -0.211 $ & $ -0.477 $ & $ -0.102 $ & 1 \\
			\midrule
			\multicolumn{8}{l}{Panel A: Correlation matrix on last day in sample}\\
			&Basic Materials	&Technology	&Healthcare	&Industrial Goods&	Services	&Consumer Goods &	Financial&	Utilities \\
			\midrule
			Basic Materials & 1     &       &       &       &       &       &       &  \\
			Technology & 0.57  & 1     &       &       &       &       &       &  \\
			Healthcare & 0.065 & $ -0.356 $ & 1     &       &       &       &       &  \\
			Industrial Goods & $ -0.163 $ & $ -0.345 $ & 0.381 & 1     &       &       &       &  \\
			Services & 0.454 & 0.661 & $ -0.526  $& $ -0.244 $ & 1     &       &       &  \\
			Consumer Goods & 0.661 & 0.719 & $ -0.45 $ & $ -0.527 $ & 0.675 & 1     &       &  \\
			Financial & 0.317 & 0.700   & $ -0.477 $ & $ -0.641 $ & 0.460  & 0.554 & 1     &  \\
			Utilities & 0.096 &$  -0.173 $ & 0.447 & 0.389 & $ -0.208 $ & $ -0.462 $ & $ -0.115 $ & 1 \\
			\bottomrule
		\end{tabular}%
	\end{adjustbox}
	\label{corr_sector_}%
\end{table}%
\end{landscape}

\pagebreak
\newpage

\setstretch{1.5}

%-----------------------------------------------------

\setstretch{1.4}
\lstset{framexleftmargin=5mm, frame=shadowbox, rulesepcolor=\color{blue}}
\begin{lstlisting}[language=json]
{"object":"Message",
"action":"create",
"data":{"id":185405776,"body":"In the last years, $PLAB&#39;s CEO bought shares at market price only once. Six months after the purchase, the stock was up by +45.24%. Check out https://www.ceo-buys.com to learn more.","created_at":"2019-12-01T00:00:00Z","user":{"id":1719889,"username":"CEOBuys","name":"CEO-Buys.com","avatar_url":"https://avatars.stocktwits.com/production/1719889/thumb-1543282592.png","avatar_url_ssl":"https://avatars.stocktwits.com/production/1719889/thumb-1543282592.png","join_date":"2018-11-19","official":false,"identity":"User","classification":[],"followers":1140,"following":0,"ideas":29818,"watchlist_stocks_count":0,"like_count":0,"plus_tier":"","premium_room":"","subscribers_count":120,"subscribed_to_count":0,"following_stocks":0,"location":"","bio":null,"website_url":null,"trading_strategy":{"assets_frequently_traded":["Equities"],"approach":"Momentum","holding_period":"Swing Trader","experience":"Professional"}},"source":{"id":4467,"title":"CEO Buys","url":"https://www.ceo-buys.com"},"symbols":[{"id":3146,"symbol":"PLAB","title":"Photronics Inc.","aliases":[],"is_following":false,"watchlist_count":683,"exchange":"NASDAQ","sector":"Technology","industry":"Semiconductor - Integrated Circuits","logo_url":"http://logos.xignite.com/NASDAQGS/00021796.gif","trending":false,"trending_score":-0.459807}],"prices":[{"id":3146,"symbol":"PLAB","price":"11.76"}],"mentioned_users":[],"entities":{"sentiment":null},"sentiment":{"sentiment_score":0.7582}},
"time":"2019-12-01T00:00:00Z"}
\end{lstlisting}
\captionof{lstlisting}{An example of JSON file for a messages on the StockTwits.}\label{listing_json_file}
\captionof*{lstlisting}{Detailed information about this format can be found in \url{https://www.json.org/json-en.html}, among others. Each JSON file has four main categories: object, action, data, and time. Except for the time of the creation for a tweet, all of the information is inside the data section. Each tweet and user has a unique id. To view the structure of the above file, simply just copy and paste it to \url{http://jsonviewer.stack.hu}.}

\setstretch{2}
\newpage\clearpage
\renewcommand{\baselinestretch}{1} % -> to reduce the inteline spacing in the references
\normalsize
\bibliographystyle{rfs}
\bibliography{Biblio_Covid19_Pandemic}

\end{document}